\documentclass[longauth]{aa}   

\usepackage{graphicx}
\usepackage{txfonts}
\usepackage{lipsum}
\usepackage{subcaption}        
\usepackage{makecell}
                                
\usepackage{lscape}             
                                
\usepackage{placeins}

\newcommand{\Ha}{\ifmmode \text{H}\alpha \else H$\alpha$\fi}
\newcommand{\nii}{\ifmmode [\text{N}\,\textsc{ii}] \else [N~{\scshape ii}]\fi}

\usepackage{amsmath}

\usepackage{xcolor}                                
\usepackage[normalem]{ulem}

\begin{document}

   \title{Star formation and morphological trends in the Antlia cluster:}

   \subtitle{Probing environmental influence out to 5$R_{200}$}


   \author{Ciria Lima-Dias
        \inst{1}, 
        Antonela Monachesi\inst{1}, 
        Gissel P. Montaguth\inst{2},
        Rodrigo F. Haack\inst{3},
        Daniela E. Olave-Rojas\inst{4},
        Hugo M\'endez-Hern\'andez\inst{1,5},
        Sergio Torres-Flores\inst{1},
        Luis Ot\'arola\inst{1},
        Juan Pablo Calder\'on\inst{3,6,7},
        Anal\'ia V. Smith Castelli \inst{3,6},
        Yara L. Jaff\'e\inst{8,5},
        Christopher P. Haines\inst{9,5},
        Diego Pallero\inst{8,5},
        Vitor M. Sampaio\inst{8,5},
        Crist\'obal Sif\'on\inst{10},
        Alvaro Alvarez-Candal\inst{11},
        Mar\'ia Argudo-Fern\'andez\inst{12,13},
        Sim\'on V\'eliz Astudillo\inst{1},
        Maiara S. Carvalho\inst{2},
        Letizia P. Cassar\`a\inst{14},
        Cheng Cheng\inst{15,16},
        Arianna Cortesi\inst{17,18},
        Ricardo Demarco\inst{19},
        Alexis Finoguenov\inst{20},
        Marcos Fonseca-Faria\inst{21},
        Thiago S. Gon\c{c}alves\inst{18},
        Eduardo Ibar\inst{5,22},
        Iván Lacerna\inst{9},
        Elismar Lösch\inst{2},
        Angela C. Krabbe\inst{2},
        Ulrike Kuchner\inst{23},
        Erik V. R. Lima \inst{2},
        Amanda R. Lopes\inst{2,6}, 
        Vitor H. Lopes-Silva\inst{18},
        Amrutha B. Manjunatha\inst{9},
        Natanael M. Cardoso\inst{24},
        Sebasti\'an Ortiz-Gómez\inst{1},
        Gustavo B. Oliveira Schwarz\inst{24},
        Umberto Rescigno\inst{9},
        Boudewijn F. Roukema\inst{25,26},
        Nicol\'as Tejos\inst{10},
        Eduardo Telles\inst{27},
        Benedetta Vulcani\inst{28}, 
        Stephane V. Werner\inst{29,30},
        Laerte Sodr\'e Jr\inst{2},
        Fábio R. Herpich\inst{31},
        Felipe Almeida-Fernandes\inst{32},
        Antonio Kanaan\inst{33},
        Tiago Ribeiro\inst{34},
        William Schoenell\inst{35},
        Claudia Mendes de Oliveira\inst{2}
        }

   \institute{Departamento de Astronom\'ia, Universidad de La Serena, Avda. Ra\'ul Bitr\'an 1305, La Serena, Chile\\
           \email{clima@userena.cl}
      \and Instituto de Astronomia, Geof\'isica e Ci\^encias Atmosf\'ericas da Universidade de S\~ao Paulo, Cidade Universit\'aria, CEP:05508-900, S\~ao Paulo, SP, Brazil
      \and Facultad de Ciencias Astron\'omicas y Geof\'isicas, Universidad Nacional de La Plata, Paseo del Bosque s/n, B1900FWA, Argentina
      \and Departamento de Tecnolog\'ias Industriales, Facultad de Ingenier\'{i}a, Universidad de Talca, Los Niches km 1, Curic\'o, Chile
      \and Millennium Nucleus for Galaxies (MINGAL), Valpara\'iso, Chile
      \and Instituto de Astrof\'isica de La Plata, CONICET-UNLP, Paseo del Bosque s/n, B1900FWA, Argentina
      \and Consejo  de Investigaciones Cient\'ificas y T\'ecnicas, Ciudad Aut\'onoma de Buenos Aires, Argentina
      \and Departamento de F\'isica, Universidad T\'ecnica Federico Santa Mar\'ia, Avenida Espa\~na 1680, Valpara\'iso, Chile
      \and Instituto de Astronom\'ia y Ciencias Planetarias (INCT), Universidad de Atacama, Copayapu 485, Copiap\'o, Chile
      \and Instituto de F\'isica, Pontificia Universidad Cat\'olica de Valpara\'iso, Casilla 4059, Valpara\'iso, Chile
      \and Instituto de Astrof\'isica de Andaluc\'ia – Consejo Superior de Investigaciones Cient\'ificas (IAA-CSIC), Glorieta de la Astronom\'ia S/N, E-18008, Granada, Spain
      \and Departamento de F\'isica Te\'orica y del Cosmos, Edificio Mecenas, Campus Fuentenueva, Universidad de Granada, E-18071 Granada, Spain
      \and Instituto Universitario Carlos I de F\'isica Te\'orica y Computacional, Universidad de Granada, 18071 Granada, Spain
      \and INAF–IASF Milano, Via A. Corti 12, 20133 Milano, Italy
      \and Chinese Academy of Sciences South America Center for Astronomy, National Astronomical Observatories, CAS, Beijing 100101, China
      \and Key Laboratory of Optical Astronomy, NAOC, 20A Datun Road, Chaoyang District, Beijing 100101, China
      \and Physics Institute, Federal University of Rio de Janeiro, Av. Athos da Silveira Ramos 149, Cidade Universit\'aria, CEP 21941-909, Rio de Janeiro, RJ, Brazil
      \and Valongo Observatory, Federal University of Rio de Janeiro, Ladeira do Pedro Antonio 43, Sa\'ude, Rio de Janeiro, RJ, Brazil, CEP 20080-090
      \and Institute of Astrophysics, Facultad de Ciencias Exactas, Universidad Andr\'es Bello, Sede Concepci\'on, Talcahuano, Chile
      \and Department of Physics, University of Helsinki, Gustaf H\"allstr\"omin katu 2, 00560 Helsinki, Finland
      \and Laborat\'orio Nacional de Astrof\'isica, Rua dos Estados Unidos, 154, CEP 37504-364, Itajub\'a, MG, Brazil
      \and Instituto de F\'isica y Astronom\'ia, Universidad de Valpara\'iso, Avda. Gran Breta\~na 1111, Valpara\'iso, Chile
      \and School of Physics and Astronomy, University of Nottingham, Nottingham NG7 2RD, UK
      \and Escola Polit\'ecnica, Universidade de S\~ao Paulo, Av. Prof. Luciano Gualberto, travessa do Politecnico, 380, S\~ao Paulo, 05508-010, Brazil
      \and Institute of Astronomy, Faculty of Physics, Astronomy and Informatics, Nicolaus Copernicus University, Grudzi\k{a}dzka 5, PL-87-100 Toru\'n, Poland
      \and Univ. Lyon1, Ens de Lyon, CNRS, Centre de Recherche Astrophysique de Lyon (CRAL) UMR5574, F-69230 Saint-Genis-Laval, France
      \and Observat\'orio Nacional - MCTI (ON), Rua Gal. Jos\'e Cristino 77, S\~ao Crist\'ov\~ao, 20921-400, Rio de Janeiro, Brazil
      \and INAF–Osservatorio Astronomico di Padova, Vicolo dell’Osservatorio 5, I-35122 Padova, Italy
      \and Centre for Extragalactic Astronomy, Department of Physics, Durham University, South Road, Durham DH1 3LE, UK
      \and Institute for Computational Cosmology, Department of Physics, Durham University, South Road, Durham DH1 3LE, UK
      \and Laboratório Nacional de Astrofísica (LNA/MCTI), Rua Estados Unidos, 154, Itajubá 37504-364, Brazil
      \and Instituto Nacional de Pesquisas Espaciais, Av. dos Astronautas 1758, Jardim da Granja,12227-010 S\~ao Jos\'e dos Campos, SP, Brazil
      \and Departamento de F\'isica, Universidade Federal de Santa Catarina, Florian\'opolis, SC, 88040-900, Brazil
      \and NOAO, 950 North Cherry Ave., Tucson, AZ 85719, United States
      \and The Observatories of the Carnegie Institution for Science, 813 Santa Barbara St, Pasadena, CA 91101, USA }

   \date{Received September 30, 20XX}

 
  \abstract
   {Galaxy evolution in dense environments, such as clusters, is strongly affected by environmental processes, which can significantly alter both morphology and star formation activity. Studying galaxies over a wide range of cluster-centric distances makes it possible to probe how these mechanisms operate across different environmental densities and to investigate pre-processing in the cluster infall regions. The Antlia cluster is particularly interesting due to its proximity and dynamical stage, but it still lacks investigation towards larger radii from the cluster core.}
   {We investigated how galaxy morphology and star formation rates vary due to environmental effects, extending this analysis out to $5R_{200}$ using Southern Photometric Local Universe Survey (S-PLUS) data.} 
   {We derived star formation rates from \Ha\, emission using the $J0660$ band of S-PLUS and the Sérsic index using \texttt{GALFITM}, for which we limit our analysis to galaxies with magnitudes brighter than $m_r = 16$. Our main analysis focuses on 154 galaxies with spectroscopic redshifts consistent with the Antlia cluster. We split galaxies between early- and late-type systems (ETGs and LTGs, respectively), using a combination of colour $(u-r)$ and Sérsic index, and between quenched and star-forming (QGs and SFGs, respectively). We further explored their environmental dependence using projected phase–space diagrams, substructure, and projected local density.}
   {We find that the Antlia cluster and its infall regions are dominated by SFGs from 1 to 5$R_{200}$, whereas within 1$R_{200}$ it exhibits a higher fraction of QGs. Early-type galaxies and LTGs show similar radial trends to those of QGs and SFGs, respectively. However, within 1$R_{200}$, Antlia exhibits approximately the same fraction of ETGs and LTGs. Regarding the local density ($\Sigma_{10}[\mathrm{Mpc}^{-2}]$), QGs dominate regions with $\log(\Sigma_{10}[\mathrm{Mpc}^{-2}])$ $\gtrsim$ 1.5, while LTGs dominate regions with $\log(\Sigma_{10}[\mathrm{Mpc}^{-2}])$ $\lesssim$ 1.0. We also identify substructures extending from the central regions to the outskirts of the cluster. Substructures outside the central region are dominated by SFGs and LTGs, with fractions similar to those of non-substructure galaxies. However, when the massive central structure is included, the SFG fraction within substructures decreases relative to galaxies not associated with any detected substructure, suggesting that the central structure hosts a more environmentally processed population.}
   {Altogether, the presence of a massive substructure in the central region ($R<1R_{200}$), together with additional substructures detected out to $5R_{200}$, indicates that the Antlia cluster is still undergoing mass assembly through group accretion. This is consistent with Antlia being a dynamically active and young system assembling its galaxy population.}


\titlerunning{The Antlia cluster environment}
\authorrunning{C. Lima-Dias et al.}
   \maketitle
   \nolinenumbers

\section{Introduction}

The evolution of galaxies in clusters is shaped by a range of environmental mechanisms, including ram-pressure stripping \citep{Gunn1972ApJ}, harassment \citep{Moore1996Natur}, mergers \citep{Lotz2011ApJ}, strangulation \citep{Larson1980ApJ}, and gas accretion \citep{Fraternali2008MNRAS,Stewart2011ApJ}. These phenomena can vary in intensity depending on the galaxy’s distance from the cluster centre, mass, dynamical state, and large-scale structure \citep{Sampaio2021MNRAS,Aldas2025A&A,Astudillo2025A&A}. To better understand the environmental influence on the physical and morphological evolution and transformation of galaxies, it is important to study their properties in different cluster regions (cluster-centric distance, substructure, local density, etc.), from the central region (R <1$R_{200}$) with high velocity dispersion out to 5$R_{200}$\footnote{where $R_{200}$ is the radius at which the mean density is two hundred times the critical density of the Universe.} from the centre.

In particular, several studies have shown that both the virialised regions of clusters \citep{Paccagnella2016ApJ,Sampaio2022MNRAS} as well as galaxies located out to 5$R_{200}$ from the cluster centre \citep{Haines2015ApJ,Olave-Rojas2018MNRAS,Liu2019,Lopes2024MNRAS,deVos2024MNRAS} exhibit higher fractions of quenched galaxies (QGs) and early-type galaxies (ETGs) than galaxies in the field. This suggests that environmental effects are already acting in the infall regions, far beyond the cluster virial radius. In addition, simulations indicate that more massive clusters accrete a higher fraction of QGs \citep{Pallero2019MNRAS,Pallero2022MNRAS}, implying that part of the quenching and morphological transformation occurs prior to cluster infall. This phenomenon, known as pre-processing, describes the transformation of galaxies within groups and filaments before they are accreted onto clusters and has been widely proposed as a key mechanism behind the enhanced fractions of the QGs and ETGs observed in cluster outskirts and infall regions \citep[e.g.][]{Zabludoff1996ApJ,Fujita2004PASJ,Bahe2013MNRAS,Haines2015ApJ,Bianconi2018MNRAS}.

Recently, \citet{Lopes2024MNRAS} studied galaxy groups falling into clusters in the local Universe, extending their analysis out to 5$R_{200}$. They find that the fraction of star-forming galaxies (SFGs) within infalling groups is lower than that of individual galaxies (i.e. not in groups) at the same cluster-centric distance. Additionally, they reported that the fraction of ETGs in groups is higher than the fraction of ETGs among isolated galaxies not associated with any infalling group, in agreement with \citet{Sampaio2024MNRAS}. This again suggests that galaxies undergo pre-processing within groups before being accreted onto the cluster. Although these studies have established the importance of pre-processing in a statistical sense, individual nearby clusters with substructures remain crucial laboratories for testing how these mechanisms affect galaxy morphology and star formation across different environments.

Antlia is particularly well suited for this purpose due to its proximity  (38.73 Mpc, z = 0.009; \citealt{Sarkar2022MNRAS}) and complex dynamical state with clear substructures. It is located at right ascension (RA; J2000) = 10:30:03.6 and declination (Dec; J2000) = -35:19:23.88 \citep{Hess2015MNRAS}, with $R_{200}$ = 887 kpc \citep{Sarkar2022MNRAS} and $M_{200}$ = 1.3$\times10^{14}$$M_{\odot}$\footnote{where $M_{200}$ is the total mass enclosed within $R_{200}$.} \citep{Sifon2025A&A}. Its X-ray distribution peaks at the bright elliptical galaxy NGC 3268 and extends towards a subgroup of galaxies centred on NGC 3258 \citep{Wong2016ApJ}. The cluster is currently accreting a new group and exhibits an elongated galaxy distribution. Using X-ray data from Suzaku, a Japanese X-ray astronomy satellite, \citet{Wong2016ApJ} describe Antlia as dynamically young, although its outskirts within $R_{200}$ appear relaxed. Here, this term refers to a system under assembly and not yet fully relaxed, as indicated by the presence of high levels of substructure \citep{Gouin2021A&A}, elongated galaxy distribution, and ongoing group accretion. Moreover, observations suggest that Antlia is dynamically younger than Virgo, Fornax, and Hydra \citep{Ferguson1990AJ,Furusho2001PASJ,Smith2008MNRAS,Wong2016ApJ,Hu2023ApJ}. Early-type galaxies, mainly dwarf spheroidal \citep{Ferguson1991AJ} and compact elliptical galaxies \citep{Caso2024A&A}, dominate the cluster \citep{Dirsch2003A&A}.

Despite all these studies, Antlia remains relatively unexplored beyond $R_{200}$, and it is still unknown how its low mass and dynamically active nature influence the physical and morphological properties of infalling galaxies. Its proximity allows galaxies to be spatially resolved, enabling a detailed morphological analysis. This makes Antlia an ideal target for studying the impact of environmental effects on galaxy transformation, i.e. the evolution of galaxies from late-type, gas-rich discs to earlier-type, gas-poor systems. Moreover, at its redshift, galaxies with \Ha\ emission can be detected using data from the Southern Photometric Local Universe Survey \citep[S-PLUS;][]{Mendes2019MNRAS}, allowing us to derive star-formation rates. This enables a combined analysis of galaxy morphology and star-formation activity. Antlia is particularly suitable for this analysis, as its dynamical state and substructures offer valuable insights into galaxy evolution.

In summary, we use the Antlia cluster as a laboratory to investigate how environmental mechanisms, acting both inside and outside the virial radius, influence galaxy evolution in nearby clusters. Given its dynamically active nature, we also compare our results with those from more dynamically relaxed clusters to explore how different environmental conditions affect galaxy evolutionary paths. Finally, we present the Antlia galaxy target catalogue for the CHileAN Cluster galaxy Evolution Survey \citep[CHANCES;][]{Haines2023Msngr,Sifon2025A&A}, from which the galaxies studied in this work were selected. Operating since 2026, CHANCES is a survey using the 4-metre Multi-Object Spectroscopic Telescope \citep[4MOST;][]{deJong2019Msngr} and aims to obtain spectra and redshifts of galaxies in and around approximately 100 massive galaxy clusters. The survey will target galaxies extending out to 5$R_{200}$ from each cluster, enabling an unprecedented study of galaxy pre-processing. The list of clusters and their properties (sky position, mass, size, and redshift) is presented in \citet{Sifon2025A&A}, and the target selection is detailed in \citet{MendezHernandez2026A&A}.

This manuscript is structured as follows. Section \ref{sec:data} describes the data used and the construction of the parent catalogue. Section~\ref{sec:CHANCES_Members} presents the methodology adopted to select the photometric Antlia member candidates and identify the spectroscopic members. It moreover describes the procedures used to derive the structural and physical properties of the galaxies. In Sect. \ref{sec:spec_Members}, we analyse the spectroscopically confirmed galaxy sample across the cluster and infall regions. Section \ref{sec:discussion} discusses the implications of our results for galaxy evolution in different environments. Finally, Sect. \ref{sec:conclusions} summarises our main conclusions. Throughout this study, we adopted a flat cosmology with $H_{0} = 69.32$ km s$^{-1}$ Mpc$^{-1}$ and $\Omega_{m} = 0.287$ \citep{Hinshaw2013ApJS}.

\section{Data} \label{sec:data}

We used imaging data from the S-PLUS DR4, consisting of 70 fields that collectively cover $132.7~\deg^{2}$ around the Antlia cluster, reaching out to $5R_{200}$ from the cluster centre. An interactive RGB mosaic showing the full S-PLUS footprint, the spectroscopic sample analysed in this work, the $1R_{200}$ and $5R_{200}$ regions, and the positions of the two central structures previously reported in the literature is available as supplementary online material\footnote{ \url{https://nmcardoso.github.io/antlia-mosaic}}. S-PLUS provides 12-band optical photometry ($u_{\mathrm{JAVA}}$, $J0378$, $J0395$, $J0410$, $J0430$, $g_{\mathrm{SDSS}}$, $J0515$, $r_{\mathrm{SDSS}}$, $J0660$, $i_{\mathrm{SDSS}}$, $J0861$, and $z_{\mathrm{SDSS}}$), enabling both structural and emission-line analyses. Each S-PLUS image covers $1.4 \times 1.4~\deg$ ($\sim$$2~\deg^{2}$), providing homogeneous 12-band data over a large area of the sky. The details of data acquisition, reduction, calibration, and global image quality (including point spread function (PSF) characterisation and depth) are presented in \citet{Herpich2024A&A}. In addition, the Antlia and the Hydra cluster have been investigated in two other studies based on S-PLUS data: the first is by \citet{Cardoso2026AJ}, who focused on emission-line galaxies in Hydra, and the second is by \citet{Vinicius2026ApJ}, which describes the S-PLUS DR5. 

In this work, we used the S-PLUS photometric catalogues provided by \citet{Haack2024MNRAS}, who reprocessed S-PLUS iDR4 images in nearby galaxy cluster regions with three dedicated \textsc{SExtractor} \citep{Bertin1996} configurations. Their RUN~1 is optimised to detect faint and/or compact sources in the vicinity of bright galaxies; RUN~2 is tuned to obtain reliable measurements for bright and extended galaxies; and RUN~3 is designed to recover very extended galaxies with complex internal structures. By combining these runs, \citet{Haack2024MNRAS} produced improved 12-band photometry for nearby clusters, mitigating both the over-deblending of SFGs and the loss of faint objects close to bright systems in the original S-PLUS DR4 catalogues.

To select galaxies in the field of view of the Antlia cluster region, the following criteria were applied to ensure reliable photometry:

\begin{itemize}

    \item $FLUX\_RADIUS\_50\_r >3$
    \item $r_{\rm AUTO} < 20.4$
    \item $e\_{r_{\rm AUTO}} < 0.2$
    \item $FLAGS\_r < 4$
    \item $FLAGS\_i < 4$
    \item $(g_{\rm AUTO} - r_{\rm AUTO}) > - 2$
    \item $(g_{\rm AUTO} - r_{\rm AUTO}) <  2.0$
    \item $CLASS\_STAR\_g < 0.35$
    \item $CLASS\_STAR\_r < 0.35$
    \item $CLASS\_STAR\_i < 0.35$
    
\end{itemize}

The parameter $FLUX\_RADIUS\_50\_r$ corresponds to the radius (in pixels) enclosing $50\%$ of the total flux in the $r$ band and is therefore a proxy for the apparent size of the source. The $r_{\rm AUTO}$ magnitude was measured within an automatic aperture defined by Kron’s first-moment algorithm, and $e\_{r_{\rm AUTO}}$ denotes the associated photometric uncertainty. 

$FLAGS\_r$ and $FLAGS\_i$ indicate potential issues in the photometric measurement, such as deblending, truncation at the image edges, or the presence of nearby bright sources. By requiring $FLAGS\_r < 4$ and $FLAGS\_i < 4$, we excluded sources affected by the most severe problems. In addition, we imposed a colour cut in $(g_{\rm AUTO} - r_{\rm AUTO})$ to remove objects with extreme or non-physical colours, which are often associated with spurious detections or unreliable photometry.

Star–galaxy separation is based on the $CLASS\_STAR$ parameter in the $g$, $r$, and $i$ bands. $CLASS\_STAR$ is a dimensionless classifier ranging from 0 to 1, where values close to 1 correspond to sources whose light profile is consistent with the local PSF (i.e. point-like objects, typically stars), and values close to 0 correspond to extended sources (e.g. galaxies). It is not a physical measurement but a statistical estimate that depends explicitly on the PSF model and on the signal-to-noise ratio (S/N) of the detection. It becomes increasingly uncertain near the detection limit or in regions where the PSF varies significantly across the field. For the S-PLUS DR4 data, the survey depth and completeness as a function of magnitude are characterised in detail by \citet{Herpich2024A&A}. In particular, their Table~2 shows that the median $r$-band depth at ${\rm S/N} = 5$ is $r \simeq 20.4$ mag, with the deepest fields reaching $r \gtrsim 21$ mag, and their Fig.~10 indicates that the catalogues remain highly complete for $r \lesssim 20.5$. We therefore restricted our analysis to $r_{\rm AUTO} < 20.4$, i.e. to magnitudes around the $5\sigma$ depth of the main survey where source detection is still very efficient and where star–galaxy separation based on $CLASS\_STAR$ is expected to be reliable.

Following \citet{Haack2024MNRAS}, who demonstrated that a threshold of $CLASS\_STAR < 0.35$ in $g$, $r$, and $i$ yields a robust sample of resolved galaxies with consistent colours and magnitudes compared to DECaLS \citep{Dey2019} photometry in nearby cluster environments, we adopted this conservative cut in all three bands to define our galaxy sample. Requiring this condition simultaneously in $g$, $r$, and $i$ bands mitigates misclassifications due to noise or local PSF variations in individual bands and ensures that our final catalogue is dominated by genuinely extended sources. All magnitudes and colours used in this work were measured using the \textsc{SExtractor} AUTO aperture.

\section{Selection of Antlia member candidates across the cluster and infall regions}
\label{sec:CHANCES_Members}

\subsection{Photometric selection of Antlia member candidates}\label{sec:CHANCES_Members_phot}

Antlia is a very low-redshift system ($z = 0.009$; \citealt{Sarkar2022MNRAS}), for which photometric redshifts are not reliable for determining cluster membership \citep[see][]{Lima2022A&C}. Thus, we adopted a specific strategy, which we describe below.

(i) \textsc{Spectroscopic cross-match.} We first cross-matched all S-PLUS sources within $5R_{200}$ of Antlia with the Southern-Hemisphere Spectroscopic Redshift Compilation\footnote{\url{https://github.com/ErikVini/SpecZCompilation}} (SHSR; \citealt{delima_specz_compilation}). This compilation contains 5\,624\,559 objects in the Southern Hemisphere (declination below $+5\deg$), including galaxies, quasi-stellar objects (QSOs), and stars. It provides spectroscopic redshifts but not the corresponding flux-calibrated spectra, which limits our ability to perform detailed spectral diagnostics. Because Antlia is a nearby cluster with many extended galaxies, we adopted a matching radius of $5\arcsec$ to minimise missed associations.

\begin{figure}
\centering
\includegraphics[width=\columnwidth]{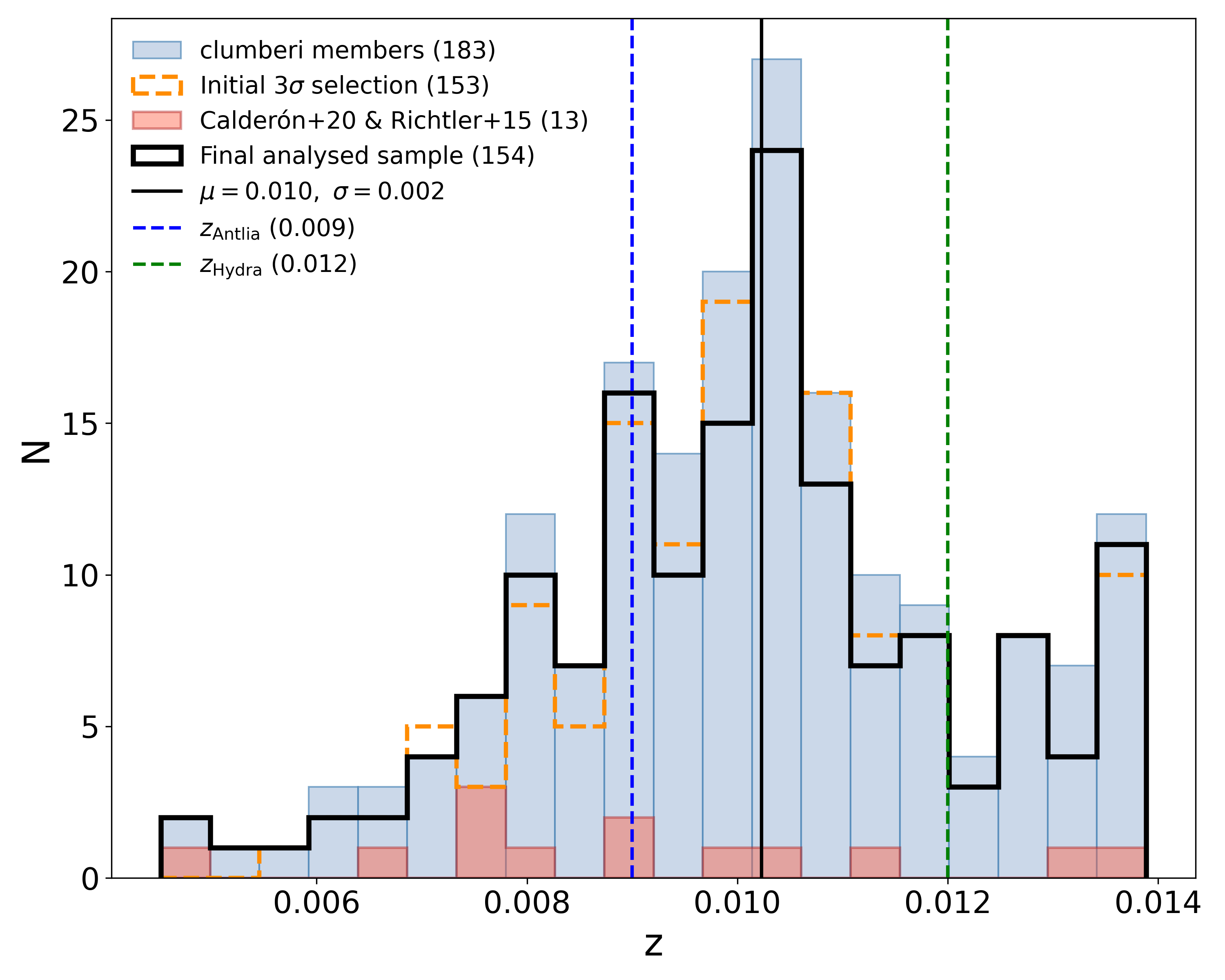}
\caption{Spectroscopic redshift distribution of galaxies within $5R_{200}$ of Antlia, showing the different membership-selection samples. The filled histogram shows the members selected with the \texttt{clumberi} algorithm, while the dashed histogram indicates the initial CHANCES selection obtained from the $3\sigma$ clipping procedure. The red histogram shows the additional literature members from \citet{Calderon2020MNRAS} and \citet{Caso2015A&A}. The black histogram outlines the final sample used in the analysis after applying the magnitude and GALFITM quality criteria. The vertical solid black line indicates the mean redshift of the initial $3\sigma$-clipped distribution, and the vertical dashed blue and green lines mark the redshifts of Antlia and Hydra, respectively. The galaxies were selected within $5R_{200}$ of Antlia, a region that includes substructures, filaments, and galaxies in the interaction region between the Antlia and Hydra clusters. This explains why the redshift distribution is not centred on the systemic redshift of Antlia.}
\label{fig:z_distribution}
\end{figure}

(ii) \textsc{Spectroscopic members.} Cluster galaxies with spectroscopic redshifts were identified by applying a $3\sigma$ clipping procedure to the redshift distribution of galaxies within $5R_{200}$. This yields a sample of 161 galaxies. The resulting redshift distribution of the selected galaxies is shown in Fig.~\ref{fig:z_distribution} as a dashed orange histogram. The high-$z$ tail of the histogram includes galaxies in the region where the Antlia and Hydra clusters (at $z=0.012$) partially overlap along the line of sight.

\begin{figure}
\centering
\includegraphics[width=\columnwidth]{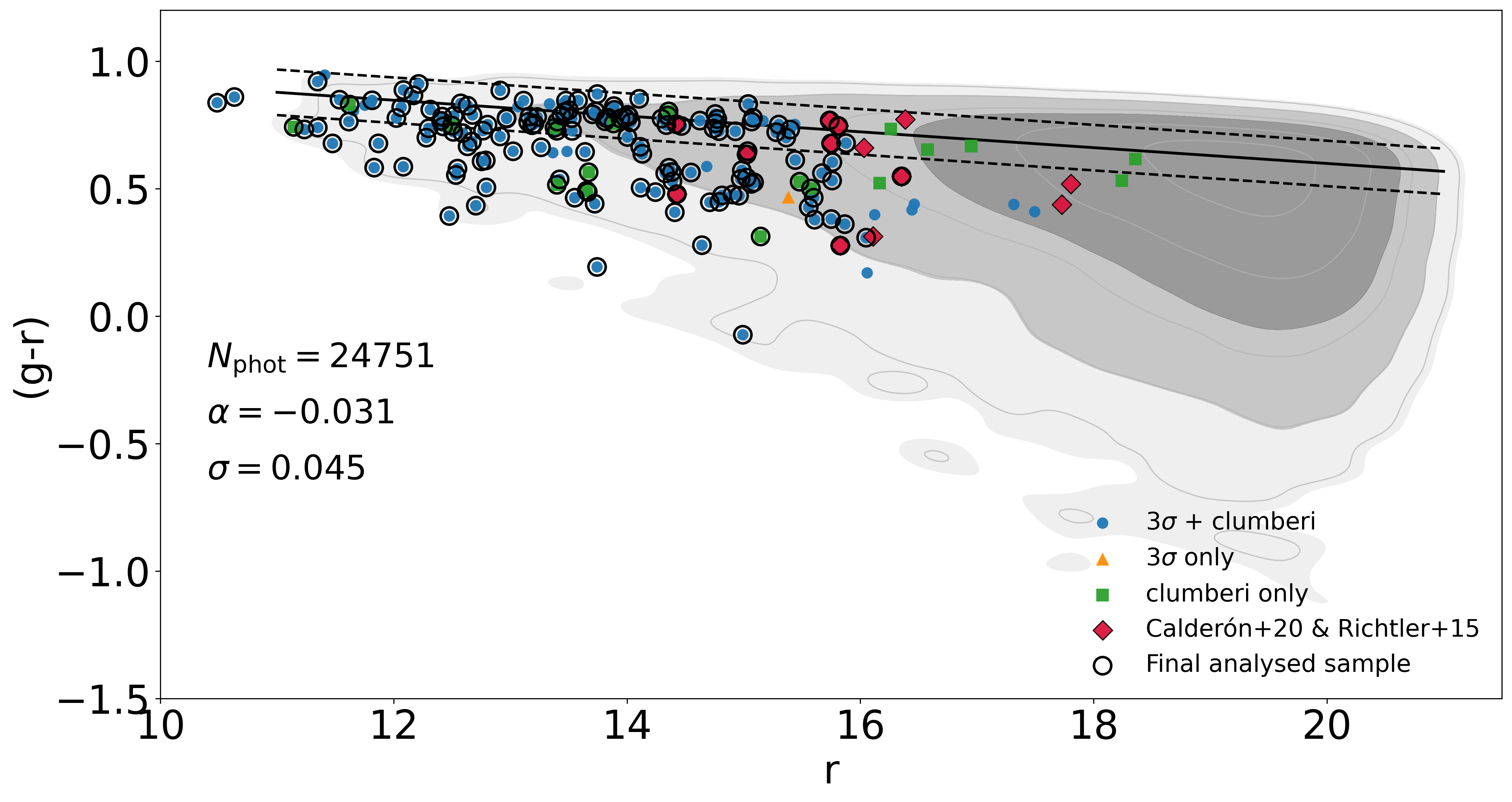}
\caption{CMD of S-PLUS galaxies within $5R_{200}$ of Antlia. The $y$-axis shows $(g-r)$, and the $x$-axis shows the apparent $r$-band magnitude. The grey-scale background shows a smoothed 2D KDE density map of the photometric candidates selected as explained in Sect. \ref{sec:CHANCES_Members_phot}. The symbols indicate the spectroscopic galaxies according to their membership-selection category: galaxies selected using both the initial $3\sigma$ clipping and \texttt{clumberi} methods (blue dots), galaxies selected using only $3\sigma$ clipping (orange triangles), galaxies selected using only \texttt{clumberi} (green squares), and additional literature members from \citet{Calderon2020MNRAS} and \citet{Caso2015A&A} (red diamonds). The open black circles highlight the final analysed sample after applying the magnitude and GALFITM quality criteria. The solid black line indicates the best-fitting red sequence, and the dashed lines correspond to the $\pm2\sigma$ envelope used to define the colour selection.
}
\label{fig:RS}
\end{figure}

(iii) \textsc{Red sequence of spectroscopic members.}
We then performed $k$ corrections\footnote{\url{https://kcorrect.readthedocs.io/en/stable/}} to the magnitudes and colours of all objects selected in ii) and constructed a colour–magnitude diagram (CMD) using $(g-r)$ versus $r$. The spectroscopically confirmed members were divided into four magnitude bins, and in each bin we fitted a double Gaussian to the colour distribution to separate the red sequence from the blue cloud. A linear fit was then applied to the red-sequence galaxies. The best-fitting relation has a slope of $-0.031$, an intercept of $1.218$, a scatter of $\sigma = 0.045$, and a coefficient of determination $R^2 = 0.71$. The resulting red sequence and its $2\sigma$ envelope are shown in Fig.~\ref{fig:RS} as blue circles and orange triangles.

(iv) \textsc{Photometric member candidates.} The red-sequence fit was used to define a broader set of candidate members from photometry. From the S-PLUS catalogue described in Sect.~\ref{sec:data}, we selected all extended sources brighter than $r_{\rm AUTO} = 20.4$ that lie within $2\sigma$ above the red sequence, together with all galaxies bluer than the red sequence. These objects are shown as a grey-scale density map in Fig.~\ref{fig:RS}. In total, this colour–magnitude selection yields 24\,751 extended sources within $5R_{200}$. This photometric catalogue will be used for target selection in the 4MOST/CHANCES survey, ensuring that the Antlia targets are extended sources with colours consistent with those of the spectroscopically confirmed cluster members, thereby increasing the probability that the observed targets are genuine members of the cluster. Further details are provided in Appendix~\ref{app:chances}. 

The \textsc{Photometric member candidates} are not used in the quantitative analysis presented in this paper, since photometric selection alone is not sufficient to robustly establish cluster membership in such a nearby system as Antlia. Throughout the rest of this work, we restrict our analysis to the spectroscopically confirmed Antlia galaxies and use the photometric candidates only to illustrate the projected spatial distribution of potential cluster members.

\subsection{Selection of spectroscopically confirmed Antlia members analysed in this work}

From the photometric catalogue described in the previous section, we retained only galaxies spectroscopically confirmed at Antlia's redshift. To test the reliability of the membership selection using solely $3\sigma$ clipping (see item ii in Sect.~\ref{sec:CHANCES_Members_phot}), we also applied the \texttt{clumberi} (CLUster MemBER Identifier) module of the \texttt{CALSAGOS} code (Clustering ALgorithmS Applied to Galaxies in Overdense Systems; \citealt{CALSAGOS2023MNRAS}). This module models the 3D distribution of galaxies in projected position and redshift with Gaussian mixture models \citep[GMMs;][]{2010muratov}, and combines this with a $3\sigma$ clipping procedure to downweight and remove outliers. In addition to the Southern Hemisphere Spectroscopic Redshift, we included 13 additional galaxies from \citet{Calderon2020MNRAS} and \citet{Caso2015A&A} that were not part of the SHSR compilation at the time the CHANCES catalogue was constructed. For these systems, we used only the measured redshifts, since the spectra are not available in a homogeneous format. We therefore did not attempt any spectral-synthesis or emission-line analysis.

Applying \texttt{clumberi} to galaxies in the Antlia cluster region, within $5R_{200}$, yields 183 members, with a line-of-sight velocity dispersion of $\sigma = 586 \pm 34$ km s$^{-1}$. The uncertainty in the velocity dispersion was estimated via bootstrap resampling. Despite differences in sample selection and radial coverage, this value is consistent with previous estimates for the Antlia cluster, which report velocity dispersions in the range $\sim 524$--$698$ km s$^{-1}$ \citep[e.g.][]{Hess2015MNRAS,Caso2015A&A}. The list of galaxies flagged by \texttt{CALSAGOS} as dynamically associated with Antlia is very similar to that obtained with the simpler $3\sigma$ clipping described in Sect.~\ref{sec:CHANCES_Members_phot}. Only one galaxy was selected only after $3\sigma$ clipping. The GMM-based approach recovers 22 additional galaxies located near the edges of the redshift distribution. This set of 183 galaxies includes both galaxies in the virialised cluster region and systems in the surrounding infall region. 

The final sample used in the analysis was obtained after applying the magnitude limit and structural-quality criteria. Specifically, as shown in Fig.~\ref{fig:RS}, the vast majority of spectroscopic members are brighter than $r_{\rm AUTO} \simeq 16$, with only a handful of fainter objects. To maintain a homogeneous sample with robust structural measurements and high-S/N photometry, we restricted the analysis to galaxies with $r_{\rm AUTO} < 16$ that also have acceptable \textsc{GALFITM} models according to the criteria described in Sect.~\ref{sec:galfitm}. This results in a final working sample of 154 galaxies.

Because our analysis relies on a heterogeneous spectroscopic compilation, we corrected all fractions for spectroscopic incompleteness as a function of both apparent magnitude and cluster-centric distance. Following the approach adopted in cluster studies such as
\citet{Jaffe2013MNRAS}, we then computed the completeness in bins of $r$-band magnitude and projected radius, defining $C(m,r) = N_{\rm spec}(m,r) / N_{\rm phot}(m,r)$, where $N_{\rm spec}$ and $N_{\rm phot}$ are the numbers of spectroscopic and photometric galaxies in each bin, respectively. Each spectroscopic galaxy was assigned a weight $w = 1/C(m,r)$ based on the bin it belongs to, and all galaxy fractions were computed as weighted sums, $\sum w_{\rm type} / \sum w$. The 1D completeness trends as a function of magnitude and cluster-centric distance are shown in
Appendix~\ref{sec:appendix_completeness}.

It is important to note that the spectroscopic completeness decreases beyond $R_{200}$, where the number of available redshifts is lower, and the spectroscopic sample is drawn from a heterogeneous literature compilation. The completeness weights therefore provide a correction for the uneven sampling as a function of magnitude and projected radius, but they do not fully remove the limitations associated with low-number statistics in the infall region. For this reason, the radial trends and fractions beyond $R_{200}$ should be interpreted as qualitative indicators of the galaxy population, rather than as precise measurements of the underlying Antlia population.

\section{Exploring the Antlia spectroscopic sample across the cluster and infall regions}
\label{sec:spec_Members}

\subsection{Modelling the galaxies' light profiles with \textsc{GALFITM}}\label{sec:galfitm}

One of the goals of this study is to assess how the environment affects the structural and physical properties of galaxies. Structural parameters were therefore derived by modelling the galaxies’ light profiles, which requires well-resolved, high-S/N images. \citet{Ortiz2026A&A} compared galaxies' structural parameters obtained from S-PLUS and DECaLS/Legacy Survey images. The two datasets are complementary: S-PLUS provides 12 optical filters, which is advantageous for multi-band \textsc{GALFITM} modelling, whereas DECaLS provides fewer broadbands but deeper and better resolved images. In the $r$ band, DECaLS reaches $r\simeq23.9$ mag, compared to $r\simeq21.18$ mag for S-PLUS at $S/N>3$, corresponding to a difference of about 2.7 mag in depth. \citet{Ortiz2026A&A} also performed the comparison for a magnitude-limited sample with $r<18.5$, finding that the S-PLUS-based structural parameters remain consistent with those derived from DECaLS within the uncertainties. In this comparison, S-PLUS-based values are slightly larger, with typical offsets of $n=0.11\pm0.01$ and $R_e=0.48\pm0.09$ kpc. These differences are small compared with the scale relevant for our analysis and do not alter the morphological interpretation. Since the final Antlia sample analysed here is restricted to $r_{\rm AUTO}<16$, our galaxies are well within the magnitude range where S-PLUS structural parameters are expected to be robust.

We modelled the surface-brightness distribution of each galaxy with a single S\'ersic profile \citep{Sersic1963}, using the multi-band fitting code \textsc{GALFITM} \citep{Peng2002AJ,Peng2010AJ,Boris2022A&A}. The code takes as input the galaxy images in all available filters together with initial guesses for the total magnitudes, effective radius, S\'ersic index ($n$), position angle, and axis ratio ($b/a$). A key advantage of \textsc{GALFITM} is that it fits all bands simultaneously, allowing the structural parameters to vary smoothly with wavelength and thereby improving the stability and accuracy of the fits \citep[e.g.][]{Vulcani2014MNRAS,Vika2014MNRAS,vika2015A&A,Kuchner2017A&A,Boris2022A&A}. The suitability of S-PLUS data for this type of analysis has been extensively tested: \citet{Lima-Dias2021MNRAS} used simulated galaxies with S-PLUS-like depth, noise, and PSF to show that \textsc{GALFITM} recovers input structural parameters with typical uncertainties of 4\%. Further applications of \textsc{GALFITM} to S-PLUS data in groups and clusters can be found in \citet{Lima-Dias2024MNRAS}, \citet{Montaguth2023MNRAS,Montaguth2025A&A,Montaguth2026ApJ}, and \citet{Ortiz2026A&A}.

To carry out the modelling in an efficient and homogeneous way, we used the \textsc{MorphoPLUS} pipeline \citep{Montaguth2026ApJ}. Given the galaxy coordinates (RA, Dec), \textsc{MorphoPLUS} automatically downloads the S-PLUS images from the survey database\footnote{\url{https://splus.cloud/}}, constructs a PSF and a mask (used to exclude contaminated pixels) for each band, generates the \textsc{GALFITM} configuration files, runs the fits, and reads the outputs. The catalogue by \citet{Haack2024MNRAS} provides the input magnitudes, positions, position angles, and other quantities used as initial guesses. In our fits, all structural parameters were allowed to vary linearly with wavelength, while the central positions were tied across bands with an offset relative to the input coordinates.

\begin{figure*}
\includegraphics[width=1\textwidth]{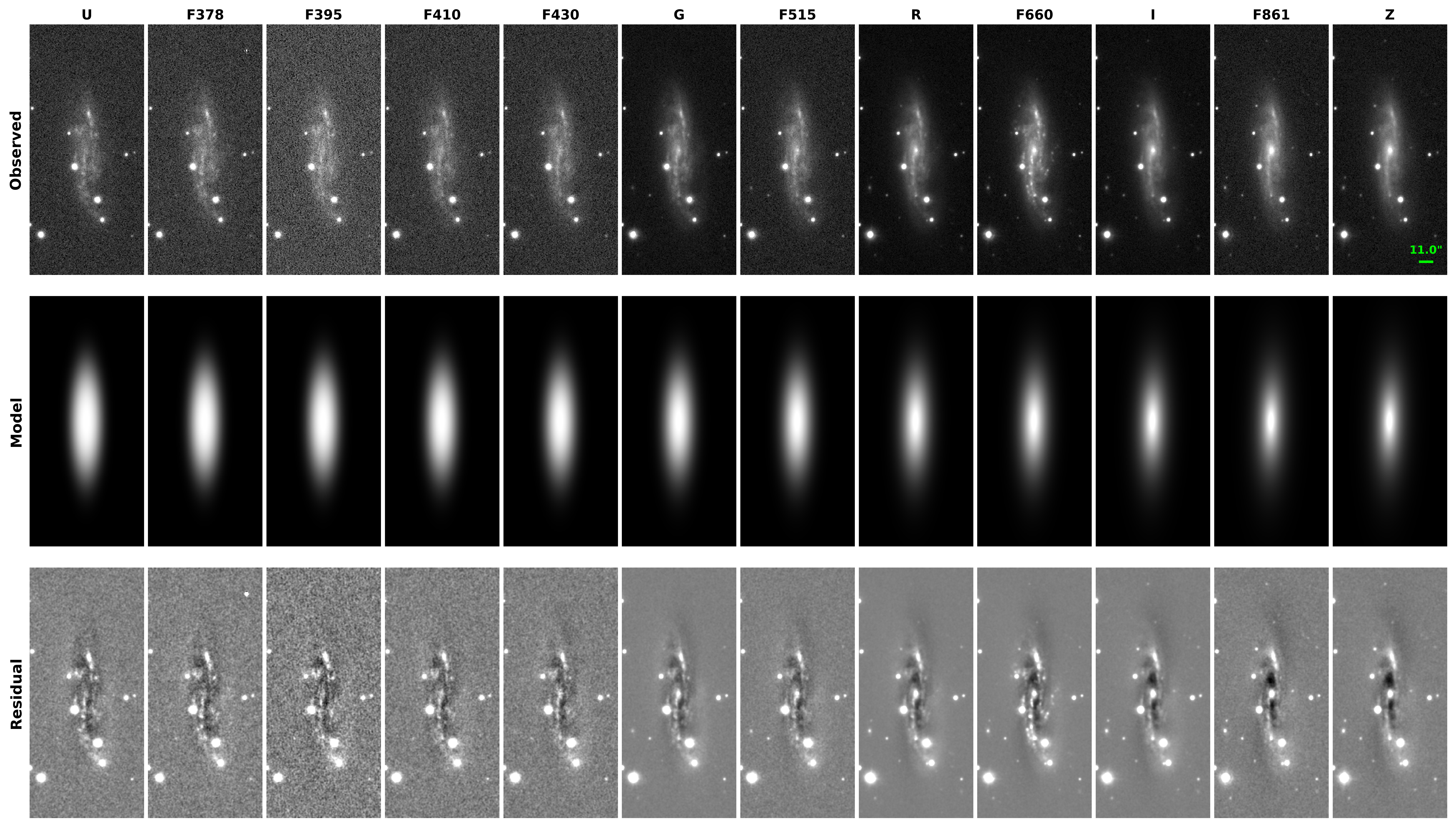}
    \caption{Galaxy located in the Antlia cluster area, as observed by S-PLUS (top panels). The models were calculated using \textsc{GALFITM} (middle panels) and the residual image (observed minus modelled; bottom panels). From left to right, the bands are $uJAVA$, $J0378$, $J0395$, $J0410$, $J0430$, $gSDSS$, $J0515$, $rSDSS$, $J0660$, $iSDSS$, $J0861$, and $zSDSS$, respectively.}
    \label{fig:galaxy_LTG}
\end{figure*}

All fitted objects were visually inspected using the data, model, and residual images. We used the reduced chi-square, $\chi^2$, only as a diagnostic to flag potentially problematic fits, not as a strict acceptance criterion. In many cases, especially for late-type systems with pronounced spiral arms or star-forming clumps, a single S\'ersic profile cannot reproduce the small-scale structure. As a consequence, the residuals increase and $\chi^2$ can reach values of a few, even when the global light distribution (total magnitude, size, ellipticity, and overall profile shape) is well modelled. Visual inspection shows that galaxies with $\chi^2 \lesssim 5$ are generally well fitted in this global sense, and we therefore adopted $\chi^2 \le 5$ as an upper limit for accepting fits in our structural analysis. This yields a final number of 154 galaxies to be analysed. Figure~\ref{fig:galaxy_LTG} shows an example of a spiral galaxy that satisfies all the acceptance criteria adopted in this work. Although spiral arms and star-forming regions remain visible in the residuals, the single-Sérsic model captures the global light distribution of the galaxy.

\subsection{Stellar mass}
\label{sec:stellar_mass}

We estimated stellar masses using the colour-mass relation of \citet{Taylor2011MNRAS}, a method previously applied to S-PLUS data \citep[e.g.][]{Lima-Dias2021MNRAS,Montaguth2023MNRAS,LopesA2025A&A}. \citet{Lima-Dias2021MNRAS} compared stellar masses derived from colour relations with masses obtained through spectral energy distribution (SED) fitting with \textsc{LePHARE} \citep{Arnouts1999MNRAS,Ilbert2006A&A} for Hydra cluster galaxies. The two methods show a strong correlation, with $R^{2}$=0.97. The median uncertainty of the colour-based masses was 0.06 dex, while the intrinsic uncertainty of the \citet{Taylor2011MNRAS} calibration is $\sim$0.1 dex. \citet{Lima-Dias2021MNRAS} also show that the \citet{Bell2003ApJS} relation gives masses that are, on average, 0.22 dex larger than those obtained with the \citet{Taylor2011MNRAS} relation, likely due to the different initial mass function (IMF) prescriptions adopted. We therefore adopted the \citet{Taylor2011MNRAS} relation in our study, propagating the S-PLUS photometric errors in $g$ and $i$ bands to estimate the stellar mass uncertainties. Possible systematics related to stellar population models, IMF, dust attenuation, and star-formation histories may affect the absolute mass scale at the ($\sim$0.1) dex level, but they should not significantly affect the relative environmental trends analysed in this work. We therefore computed stellar masses as
\begin{equation}
\log(M_{\star}/M_{\odot}) = 1.15 + 0.70 \times (g-i) - 0.40 \times M_{i},
\end{equation}

where $M_{i}$ is the rest-frame absolute magnitude in the $i$ band. To derive $M_i$, we assumed a common distance for all galaxies equal to the distance of the Antlia cluster. The uncertainties in the stellar masses were estimated by propagating the S-PLUS photometric uncertainties in $g$ and $i$ through the \citet{Taylor2011MNRAS} relation using the Python package \textsc{uncertainties}. The resulting error on $\log M_\star$ therefore reflects the propagation of the photometric uncertainties. In addition, the intrinsic scatter of the Taylor et al. (2011) calibration is $\lesssim 0.1$ dex. Stellar masses span the range $\log(M_\star/M_\odot) \approx 8.0$--$11.1$, with a median value of $\log(M_\star/M_\odot) \approx 9.7$ (see Appendix~\ref{sec:mass_appendix}).

\subsection{Star formation rate}
\label{sec:sfr}

One of the main advantages of S-PLUS is the presence of narrow-band filters, in particular $J0660$. For nearby galaxies ($z \lesssim 0.019$), $J0660$ covers the \Ha\, emission line ($\lambda = 6562.8$\,\AA), allowing us to estimate the combined \Ha+\nii\, ($\lambda\lambda 6548,6583$\,\AA) emission using the 3-Filters Method \citep[3FM;][]{Pascual2007PASP}. In this technique, two broadband filters are used to trace the continuum around the line, while a narrow band isolates the line complex. In our case, the broadbands $r$ and $i$ provide the local continuum, and $J0660$ samples the \Ha+\nii\, emission. The 3FM has already been successfully applied to S-PLUS \citep{Lima-Dias2021MNRAS,Gondhalekar2024MNRAS,Grossi2025ApJ,LopesA2025A&A} and S-PLUS-like data \citep[J-PLUS;][]{Cenarro2014SPIE,Vilella-Rojo2015A&A}. Further details of our implementation are given in Appendix~\ref{sec:3FM}.

For each galaxy, the 3FM yields an observed line flux $F_{\Ha+\nii}$ that includes both \Ha\ and \nii. To recover the pure \Ha\ flux, we applied the empirical, colour-dependent correction derived by \citet{Vilella-Rojo2015A&A}, which relates
$F_{\Ha}$ and $F_{\Ha+\nii}$ as

\begin{equation}
    \log F_{\Ha}= 
    \begin{cases}
    0.989\,\log F_{\Ha+\nii} - 0.193, & \text{if } (g-i) \leq 0.5 \\
    0.954\,\log F_{\Ha+\nii} - 0.753, & \text{if } (g-i) > 0.5 \; ,
	\end{cases}
	\label{eq:NII_cor}
\end{equation}

where $F_{\Ha}$ is the flux of the \Ha\ line and $F_{\Ha+\nii}$ is the combined \Ha\,+ \nii\ flux. We then converted $F_{\Ha}$ into an \Ha\ luminosity and derived the dust-uncorrected star formation rate, $\mathrm{SFR}_{\mathrm{obs}}(\Ha)$, using the classical calibration by \citet{Kennicutt1998}.

To estimate the intrinsic star formation rate, $\mathrm{SFR}_{\mathrm{int}}$, we corrected for dust attenuation following \citet{Ly2007ApJ}, who relates the observed and intrinsic SFRs via

\begin{equation}
\begin{split}
   \log \bigl(\mathrm{SFR}_{\mathrm{obs}}(\Ha)\bigr)
   = \log \bigl(\mathrm{SFR}_{\mathrm{int}}\bigr) - 2.360 \\
   \times \log\left [ \frac{0.797\,\log(\mathrm{SFR}_{\mathrm{int}})+3.786}{2.86} \right ] \; .
	\label{eq:SFR_corre}
\end{split}
\end{equation}

This equation was solved numerically for $\mathrm{SFR}_{\mathrm{int}}$, and the specific star formation rate (sSFR) was then obtained as
$\mathrm{sSFR} = \mathrm{SFR}_{\mathrm{int}}/M_\star$, using the stellar masses derived in Sect.~\ref{sec:stellar_mass}. Following
\citet{Wetzel2013,Wetzel2014}, galaxies with $\log(\mathrm{sSFR}\,[\mathrm{yr}^{-1}]) > -11$ were classified as SFGs, while those below this threshold were considered QGs.

The SFRs derived in this work should be interpreted as homogeneous photometric H$\alpha$-based estimates obtained with the 3FM. This methodology has been tested for the J-PLUS photometric system by \citet{Vilella-Rojo2015A&A} as follows. Using simulated J-PLUS photometry from SDSS spectra of SFGs, they showed that the 3FM recovers the \Ha\,+ \nii\ flux with a mean underestimation of approximately 9\% relative to the spectroscopic values. Therefore, while the method provides a robust and homogeneous way to estimate recent star formation over wide areas, small systematic offsets in the absolute SFR scale are expected. Additionally, \citet{LopesA2025A&A} compared H$\alpha$+[NII] maps obtained from S-PLUS images with H$\alpha$ maps from MUSE/F3D for Fornax galaxies and showed that the method can recover line emission down to fluxes of approximately $2\times10^{-17},{\rm erg,s^{-1},cm^{-2}}$. This comparison also highlights the main limitations relevant to our analysis: the narrow-band flux includes \nii\, diffuse low-surface-brightness emission can be difficult to recover, and the results may depend on the depth of the individual S-PLUS fields and on uncertainties in the continuum subtraction. For these reasons, we did not use the SFRs to derive precise star-forming main-sequence parameters. Instead, we used them primarily to separate galaxies with significant ongoing star formation from systems with little or no current star formation.

\begin{figure}
\centering
\includegraphics[width=\columnwidth]{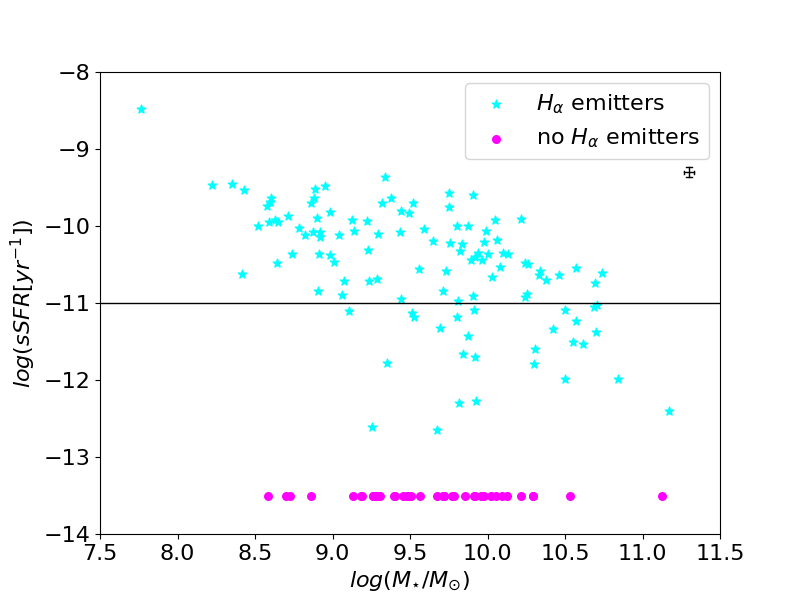}
    \caption{sSFR as a function of stellar mass. The cyan stars indicate galaxies exhibiting \Ha\ emission, whereas the magenta circles represent galaxies without detectable \Ha\ emission. The black error bar denotes the mean uncertainty in the sSFR for the entire sample. The horizontal black line marks the threshold separating SFGs from the QGs population \citep{Wetzel2013}.}
    \label{fig:sSFR_Mass}
\end{figure}

Figure~\ref{fig:sSFR_Mass} shows the relation between sSFR and stellar mass. Galaxies with \Ha\ detections are plotted as cyan stars, whereas magenta circles indicate systems without detectable \Ha\ emission. The horizontal black line marks the adopted division between SFGs and QGs, and the black error bar denotes the mean uncertainty in sSFR for the sample. We note that the stellar masses were estimated using the $(g-i)$ colour and $i$-band luminosity, while the SFRs were derived from the $r$, $i$, and $J0660$ bands. Therefore, part of the same broadband photometric information enters both estimates. This can introduce correlated scatter in the SFR-$M_{\star}$ plane. Nevertheless, our analysis does not rely on fitting the slope or normalisation of the star-forming main sequence. The SFR-$M_{\star}$ plane is used only as a diagnostic, and the adopted SFG/QG classification is based on a broad sSFR threshold together with the presence or absence of significant \Ha\ emission. We therefore expect the main separation between star-forming and quiescent galaxies to be robust against small correlated uncertainties, although absolute SFR and sSFR values should be interpreted with this limitation in mind.\\

\subsection{Classification as early- and late-type galaxies}
\label{sec:morph_class}

We used the $(u-r)$ colour and the S\'ersic index in the $r$ band, $n_r$, to separate galaxies into ETG and LTG systems. In this plane, ETGs are generally characterised by higher S\'ersic indices (more bulge-dominated systems) and redder colours, while LTGs tend to have lower $n_r$ (more disc-dominated systems) and bluer colours. Following \citet{vika2015A&A}, we adopted fixed thresholds of $n_r = 2.5$ and $(u-r) = 2.3$ to define four quadrants in the $(u-r)$–$n_r$ diagram: galaxies with $n_r \ge 2.5$ and $(u-r) \ge 2.3$ are classified as ETGs, while systems with $n_r < 2.5$ and $(u-r) < 2.3$ are classified as LTGs.

\begin{figure}
\centering
\includegraphics[width=\columnwidth]{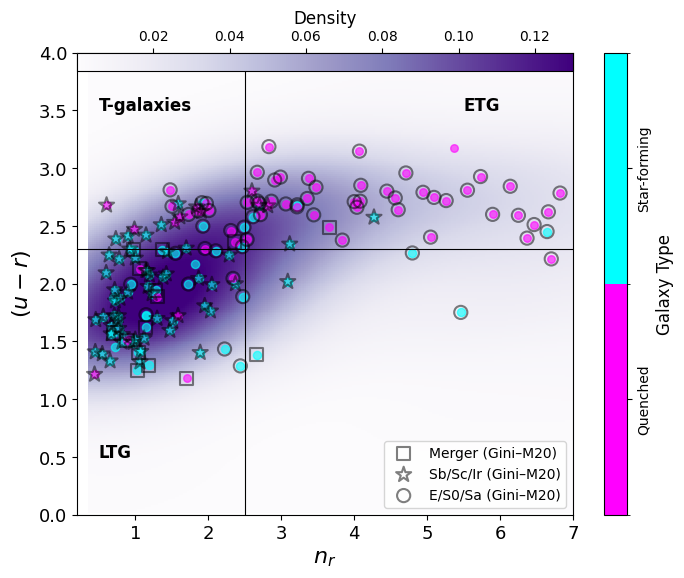}
    \caption{Distribution of galaxies in the $(u-r)$ colour vs $r$-band S\'ersic index, $n_r$. The coloured symbols indicate the spectroscopic working sample (Sect.~\ref{sec:spec_Members}), with QGs in magenta and SFGs in cyan. The vertical and horizontal lines mark $n_r = 2.5$ and $(u-r) = 2.3$, respectively, and are used to define broad ETG/LTG and transition regions in this diagram. The symbol shapes further encode the non-parametric classification from the Gini–$M_{20}$ diagram. The open squares indicate mergers, the stars indicate Sb/Sc/Irr systems, and the circles indicate E/S0/Sa galaxies, as shown in the legend. The background density map shows all CHANCES Low-$z$ Bright galaxies (see Appendix~\ref {app:chances}) within $5R_{200}$ of Antlia that are brighter than $r_{\rm AUTO}=16$ and have acceptable single-S\'ersic fits (Sect.~\ref{sec:galfitm}). }
    \label{fig:vika_density}
\end{figure}

Figure~\ref{fig:vika_density} shows the $(u-r)$ versus $n_r$ plane for the spectroscopic sample used in this work, with QGs shown in magenta and SFGs in cyan. As expected, most QGs populate the region associated with ETGs, whereas SFGs are predominantly found in the LTG region. In the same figure, the underlying density map represents the CHANCES Low-$z$ Bright photometric catalogue in the Antlia area (see Appendix \ref{app:chances}), restricted to $r_{\rm AUTO}<16$. The fact that the underlying photometric catalogue follows the distribution traced by the spectroscopic sample supports the reliability of the photometric selection to identify probable cluster members.

Using the above thresholds, we find that LTGs account for 52$\pm$5\% of the spectroscopic sample, while ETGs represent 29$\pm$4\%. These fractions refer to galaxies within $5R_{200}$ and should be interpreted in the context of the magnitude and spectroscopic-selection limits discussed previously. A third region in the diagram, occupied by galaxies with disc-like S\'ersic indices ($n_r < 2.5$) but relatively red colours ($(u-r) \ge 2.3$), corresponds to the so-called transition or T-galaxies \citep{Montaguth2023MNRAS}. These transition galaxies are not analysed in detail here and will be explored in a forthcoming work. Their behaviour in galaxy groups has been studied by \citet{Montaguth2023MNRAS}. The ETG and LTG fractions quoted above for the full sample were computed relative to the entire spectroscopic sample; consequently, they do not sum to 100\%, because of the fraction of galaxies that lie in the first ($\sim$17\%) and fourth regions ($\sim$4\%) of the $(u-r)$–$n_r$ diagram (Fig.~\ref{fig:vika_density}).

As an independent check of our morphology classification, we also measured a set of non-parametric indices from the S-PLUS $r$-band images using the \textsc{galmex} code \citep{Sampaio2026A&A}. These include the CAS parameters (concentration $C$, asymmetry $A$, and smoothness $S$) and the MEGG suite ($M_{20}$, entropy $E$, Gini coefficient $G$, and gradient pattern asymmetry $G2$). As shown in Appendix~\ref{sec:nonparam_appendix}, the distributions of these indices across the $(u-r)$-$n_r$ plane, as well as the location of Antlia galaxies in the Gini-$M_{20}$ diagram relative to the classification boundaries of \citet{Lotz2008ApJ}, are broadly consistent with the ETG/LTG regions defined by our simple $(u-r, n_r)$ cuts. Figure~\ref{fig:vika_density} shows the morphological classes E/S0/Sa, Sb/Sc/Ir, and Merger, using different symbols. This supports the interpretation that the parametric and non-parametric diagnostics are tracing the same underlying morphological bimodality.

\subsection{Spatial distribution of ETGs, LTGs, QGs, and SFGs}

The impact of environment on galaxy evolution depends not only on the global properties of the host cluster (e.g. its dynamical state and mass) but also on the galaxy’s cluster-centric distance and on the local density of neighbouring systems \citep[e.g.][]{Bahe2013MNRAS,Lopez-Gutierrez2022MNRAS}. The latter is particularly relevant for galaxies that may have been accreted as part of groups, where pre-processing and galaxy–galaxy interactions can be important. In this section we explore how the fractions of quenched and SFGs, as well as of early- and late-type systems, vary with projected cluster-centric radius and local density.

\begin{figure}
\centering
\includegraphics[width=\columnwidth]{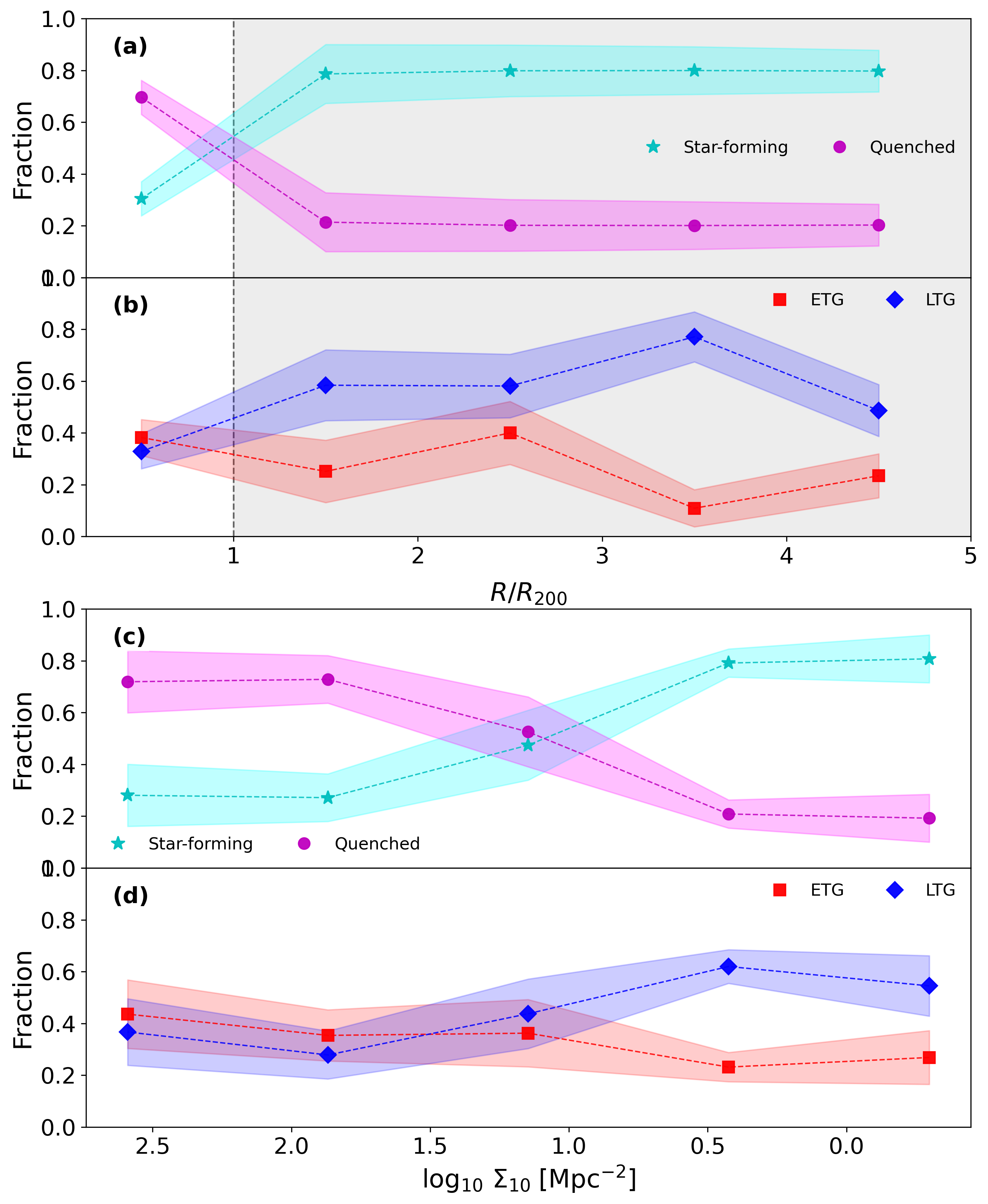}
\caption{Completeness-corrected fractions of QGs (magenta circles) and SFGs (cyan stars) (panel a), and of ETGs (red squares) and LTGs (blue diamonds) galaxies (panel b), as a function of projected cluster-centric distance in units of $R_{200}$. Panels c and d: Fractions, as a function of the projected local density $\log_{10}\Sigma_{10}\,[\mathrm{Mpc}^{-2}]$. In all panels the analysis is restricted to galaxies with $r_{\rm AUTO}<16$. The shaded regions indicate 1$\sigma$ uncertainties, computed as $\sigma = \sqrt{p(1-p)/N}$, where $p$ is the fraction and $N$ is the number of galaxies in each bin. In panels (a) and (b), the grey-shaded region marks the radial range beyond $R_{200}$, where the spectroscopic completeness is less than 0.2 and the completeness corrections are more uncertain.
}
\label{fig:frac_Q_SF_LT_ET}
\end{figure}

Figure~\ref{fig:frac_Q_SF_LT_ET} shows the resulting completeness-corrected fractions as a function of projected cluster-centric distance in units of $R_{200}$. Panel a displays the fractions of QGs and SFGs, while panel b shows the fractions of ETGs and LTGs. Within $\sim 1\,R_{200}$, QGs dominate over SFGs, as expected in the dense central regions of a nearby cluster. Beyond this radius, the fraction of SFGs increases and becomes larger than that of QGs, consistent with the idea that the infall region contains a higher proportion of star-forming systems. In terms of morphology, the inner $R_{200}$ exhibits comparable fractions of ETGs and LTGs, whereas at larger radii the sample is dominated by LTGs.

These radial trends should be interpreted with caution. As discussed in Sect.~3.2 and shown in Appendix~C, the spectroscopic completeness decreases substantially beyond $R_{200}$, where the number of available redshifts is low and the completeness corrections are more uncertain. In addition, because spectroscopic redshifts are generally easier to obtain for galaxies with strong emission lines, the spectroscopic sample in the outskirts may be biased towards emission-line (star-forming) systems. Therefore, the high SFG fractions observed between $1R_{200}$ and $5R_{200}$ should be interpreted cautiously, as they may partly reflect the selection function of the available spectroscopic sample.

To characterise the environment on smaller scales we estimated, for each galaxy in the spectroscopic sample, the projected local number density $\Sigma_{10}$ as the surface density of the ten nearest spectroscopic neighbours, $\Sigma_{10} = 10 / A_{10}$, where $A_{10} = \pi R_{10}^2$ is the area of the circle of radius $R_{10}$ enclosing the ten closest galaxies in projection \citep{Fasano2015}. Following the discussion by \citet{Muldrew012MNRAS}, different environment estimators are sensitive to different physical scales. The $\Sigma_{10}$ should
therefore be interpreted as a relative, nearest-neighbour-based measure of local environment within our spectroscopic sample, rather than an absolute galaxy number density. Because it is computed from an incomplete, magnitude-limited spectroscopic catalogue, $\Sigma_{10}$ systematically underestimates the true density. Nevertheless, it provides a useful internal ranking of environmental density, allowing us to compare the locations of QGs and SFGs, and of ETGs and LTGs, within the same dataset.

Figure~\ref{fig:frac_Q_SF_LT_ET} presents the fractions of QGs and SFGs (panel c) and of ETGs and LTGs (panel d) as a function of $\log(\Sigma_{10})$. At projected local densities $\log(\Sigma_{10} [\mathrm{Mpc}^{-2}]) \gtrsim 1.5$, QGs become more frequent than SFGs, whereas at lower densities SFGs dominate, as expected from the classical star-formation–density relation. Morphologically, the ETG fraction increases with $\Sigma_{10}$ and becomes comparable to or higher than the LTG fraction for $\log(\Sigma_{10}) \gtrsim 1.0$, while at lower densities LTGs dominate. The error bars in both figures are uncertainties estimated as $\sigma = \sqrt{p(1-p)/N}$, where $p$ is the fraction and $N$ the number of galaxies in each bin. Overall, despite the limitations imposed by spectroscopic incompleteness and the bright magnitude cut, the observed trends are broadly consistent with the well-established morphology–density and star-formation–density relations \citep[e.g.][]{Fasano2015,Vulcani2023ApJ}.

Since projected local density and cluster-centric distance are not independent, we also examined their relation in Appendix~\ref{sec:density_radius}. We find that the highest values of $\log_{10}\Sigma_{10}$ are mainly located within $R_{200}$, while galaxies beyond $R_{200}$ generally occupy lower-density environments. Therefore, part of the trend seen in Fig.~\ref{fig:frac_Q_SF_LT_ET} as a function of $\Sigma_{10}$ reflects the underlying radial segregation of galaxy populations. Nevertheless, $\Sigma_{10}$ remains useful as a complementary environmental indicator, since it traces local density variations within the spectroscopic sample that are not captured by cluster-centric distance alone.

\subsection{Types of galaxies in the cluster substructures}

\begin{figure}
    \centering
    \includegraphics[width=0.9\columnwidth]{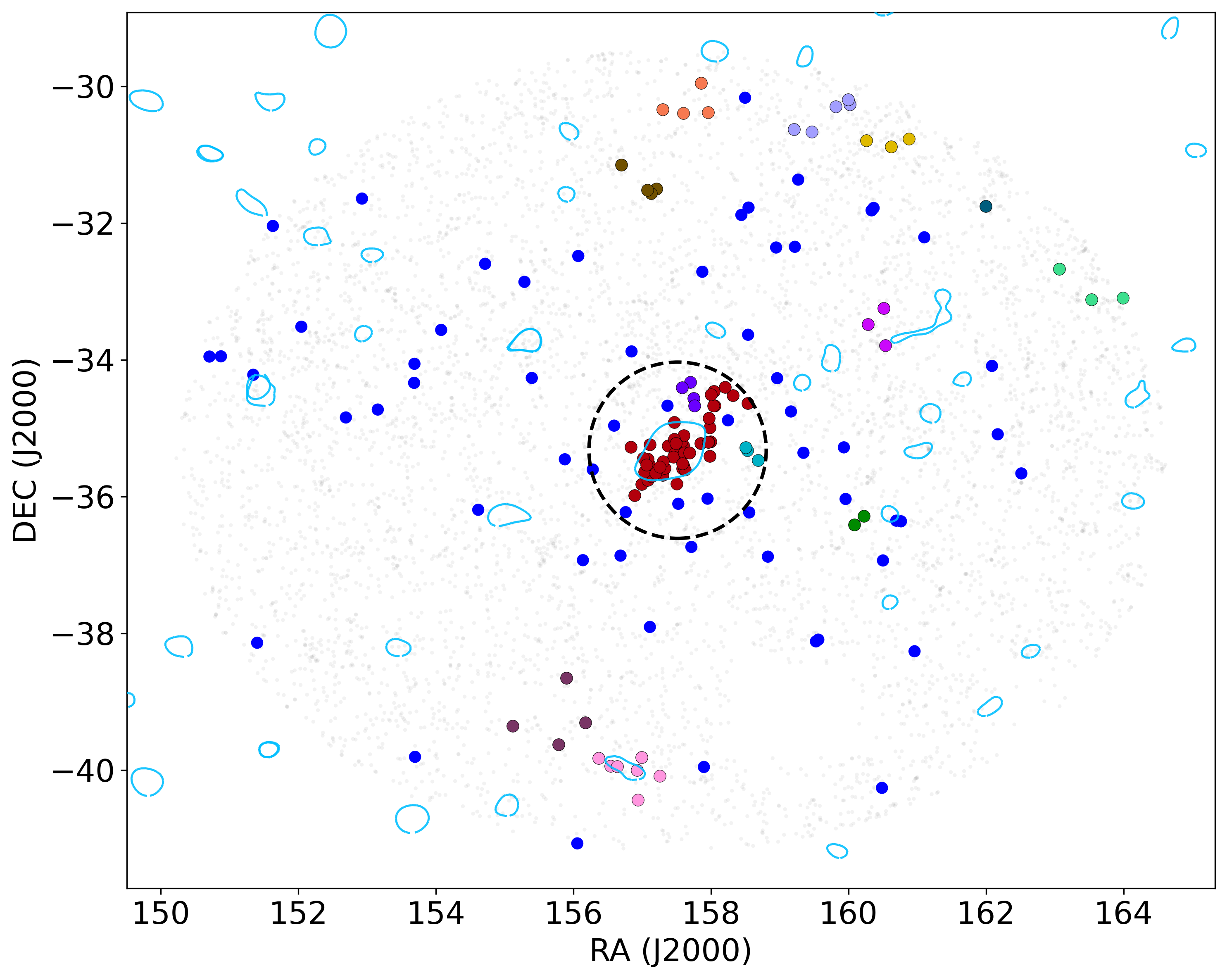}
    \caption{Spatial distribution of galaxies with redshifts consistent with the Antlia cluster, highlighting the systems identified as substructures in different colours. The dashed circle marks $1R_{200}$. The grey dots show all galaxies in the CHANCES Low-$z$ Bright catalogue (see Appendix \ref{app:chances}). The blue points denote spectroscopic galaxies that are not associated with any detected substructure. The light-blue contours trace the X-ray emission at a single surface-brightness level. Some galaxies that belong to substructures are not shown because they were excluded from the structural analysis owing to poor \textsc{GALFITM} fits. }
    \label{fig:substructure_Daniela}
\end{figure}

To identify substructures, we again used the \texttt{CALSAGOS} code, which applies the DBSCAN clustering algorithm \citep[Density-Based Spatial Clustering of Applications with Noise;][]{ester} to the projected positions of galaxies, which are shown in Fig. \ref{fig:substructure_Daniela}  for our sample of galaxies. A substructure is defined as an association of at least three galaxies. DBSCAN requires as input a distance parameter ($\varepsilon$), which specifies the search radius around each galaxy to determine whether it meets the minimum number of neighbours required to belong to a substructure. We define $\varepsilon$ as the median projected distance to the third-nearest neighbour for each galaxy.

Because the spectroscopic sampling declines towards the outskirts (see Fig. \ref{fig:substructure_Daniela}), a single global value of $\varepsilon$ would artificially enhance the detection of spurious structures at large radii. To mitigate this, we adopted a two-step approach: for identifying substructures within $3R_{200}$, $\varepsilon$ was estimated using only galaxies inside this radius, whereas for substructures beyond $3R_{200}$, $\varepsilon$ was computed considering all galaxies in the sample. 
Figure~\ref{fig:substructure_Daniela} shows the spatial distribution of the spectroscopic galaxies, highlighting the identified substructures in different colours. We also overlay a light-blue eROSITA \citep{Merloni2024A&A} contour tracing the X-ray emission, corresponding to a count rate of $10^{-6}$ counts s$^{-1}$ pixel$^{-1}$ in the 0.6--2.3 keV band, with a pixel size of 4\arcsec.

Galaxies may undergo `pre-processing' within small-scale structures and groups before being accreted onto the main cluster. Analysing the distribution of different galaxy types within these substructures is therefore of particular interest. We identify substructures at different projected distances from the cluster centre. Most are located between 4 and $5\,R_{200}$, but a massive structure (shown in red in Fig.~\ref{fig:substructure_Daniela}) is also detected in the central cluster region.

Table~\ref{tab:galaxy_types} summarises the fractions of  SFG, QG, LTG, and ETG galaxies in three environments: (i) galaxies in substructures excluding the central massive structure; (ii) galaxies in all substructures, including the central one; and (iii) galaxies not associated with any identified substructure. We find that all substructures, excluding the central one, are dominated by SFGs and LTGs. These results are discussed in more detail in Sect.~\ref{sec:substructures_D}.

\begin{table}
\caption{Galaxy types in different Antlia environments}
\label{tab:galaxy_types}
\centering
\setlength{\tabcolsep}{4pt}
\begin{tabular}{p{3.cm}cccc}
\hline\hline
Galaxy \\ environment & SFGs & QGs & LTGs & ETGs \\
\hline
\makecell{i) Substructures  \\ (excluding central)} & $76\pm7$\% & $24\pm7$\% & $60\pm8$\% & $22\pm7$\% \\
\makecell{ii) Substructures \\ (including central)} & $53\pm6$\% & $47\pm6$\% & $46\pm6$\% & $31\pm5$\% \\
\makecell{iii) Not in any \\ substructure}           & $78\pm6$\% & $22\pm6$\% & $60\pm7$\% & $26\pm6$\% \\
\hline
\end{tabular}
\tablefoot{Fractions are completeness-corrected using weights $w$. Uncertainties correspond to $1\sigma$ approximate errors computed as $\sigma_p=\sqrt{p(1-p)/N_{\rm eff}}$, with $N_{\rm eff}=(\sum w)^2/\sum w^2$.}
\end{table}

As an additional check, we compared the DBSCAN detections with the Dressler-Shectman (DS) test \citep{Dressler1988AJ}, which identifies galaxies with local kinematic deviations relative to the global cluster population. The two methods are complementary and are not expected to return identical memberships: DBSCAN identifies projected spatial overdensities, whereas the DS test uses local velocity information. We find that 7 of the 13 DBSCAN-detected substructures contain galaxies with high DS-test significance, defined as $\delta_i>\delta_{\rm lim}$, with $\delta_{\rm lim}=3\sigma_\delta$ following \citet{Girardi1997ApJ}. This agreement supports the presence of several of the substructures identified by DBSCAN, while the remaining differences are expected given the different information used by the two methods and the limitations imposed by the heterogeneous and incomplete spectroscopic sampling, especially in the cluster outskirts.

We also tested the stability of the DBSCAN group identification using the adjusted Rand index (ARI) and the Jaccard index over 100 realisations. These tests included cases in which 20\% of the galaxies were randomly removed, 80\% additional galaxies were randomly added to mimic a more complete catalogue, and galaxy positions and redshifts were resampled. The results indicate that the substructure identification is sensitive to catalogue completeness, meaning that some real substructures may be missed in the present data. However, the DBSCAN-detected substructures are stable under these tests, supporting their interpretation as robust projected associations within the limitations of the current spectroscopic sample.

\subsection{Projected phase–space diagram: constraints on accretion histories} \label{sec:ps_section}

Galaxies within and around a cluster are subject to its gravitational potential, and the environmental processes they experience depend on their cluster-centric distance, orbital velocity, and the mass of the host halo, among other properties \citep{Treu2003ApJ,Zhang2013MNRAS,Jaffe2015,Pallero2019MNRAS,Vulcani2023ApJ}. The projected phase–space (PPS) diagram is a useful tool for studying these effects: it shows the line-of-sight velocity of galaxies with respect to the cluster systemic velocity, normalised by the velocity dispersion ($|\Delta v|/\sigma$), as a function of projected cluster-centric distance normalised by $R_{200}$.

Cosmological simulations indicate that different regions of PPS are populated, on average, by galaxies with different times since first infall into the cluster and different levels of tidal mass loss, although the mapping is highly probabilistic \citep[e.g.][]{Rhee2017ApJ,Pasquali2019MNRAS}. These studies also show that the association between PPS location is broad and overlapping: virialised, infalling, and backsplash populations all contribute to each region of the diagram. Projected phase–space 'zones' should therefore be interpreted only in a statistical sense, as regions where a given class is more likely, and not as hard boundaries for individual galaxies. In this work we adopt the simple division between virialised region and a non-virialised region proposed by \citet{Mahajan2011MNRAS}.

\begin{figure*}
\centering
\includegraphics[width=\textwidth]{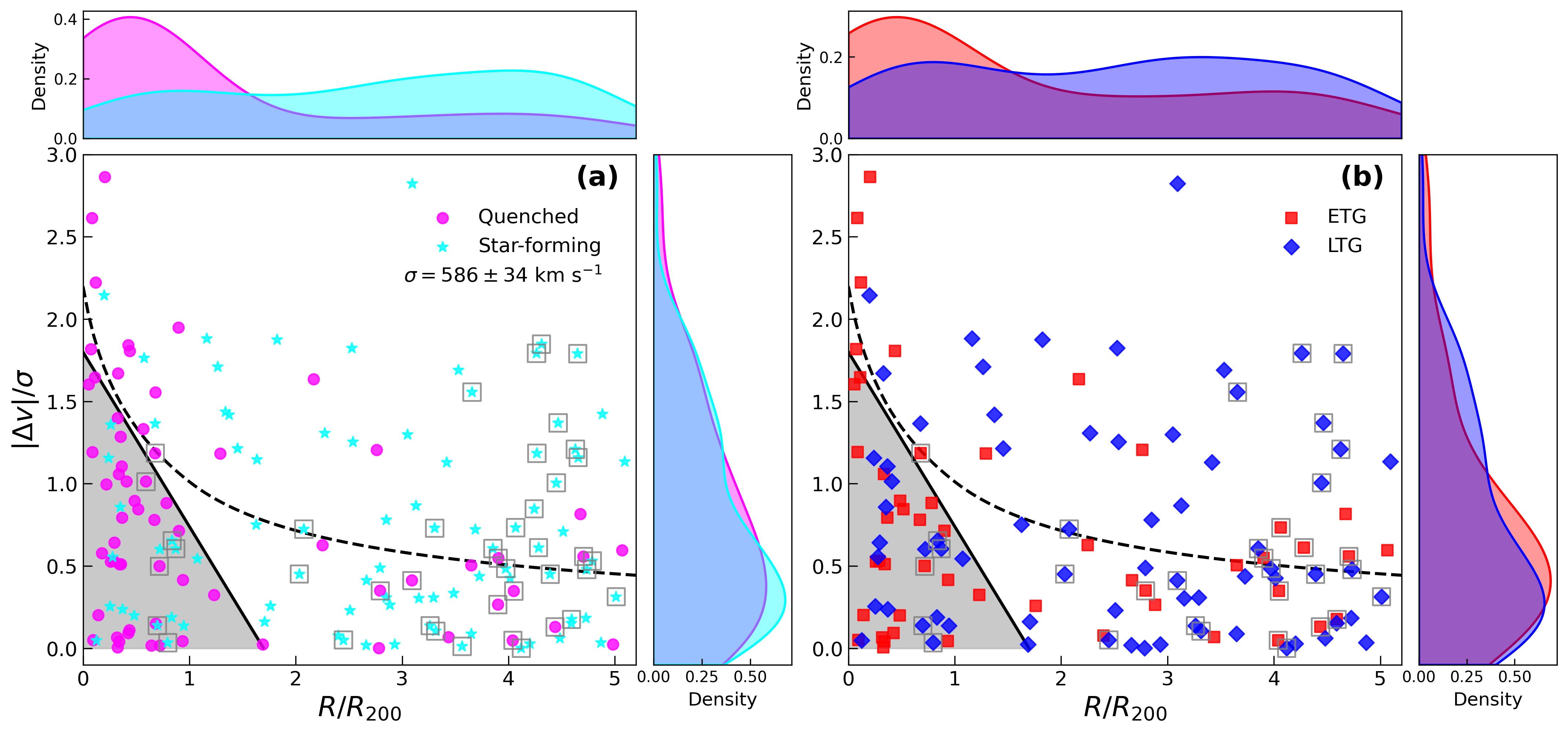}
\caption{PPS diagrams for galaxies with redshifts consistent with the Antlia cluster. The x-axis shows the projected distance from the cluster centre, normalised by $R_{200}$, while the y-axis shows the absolute line-of-sight velocity offset, normalised by the cluster velocity dispersion. Panel (a): Galaxies classified according to their star-formation activity, with QGs shown as magenta circles and SFGs as cyan stars. Panel (b): Galaxies classified according to morphology, with ETGs and LTGs shown as red squares and blue diamonds, respectively. The dashed black curve shows the escape velocity for an NFW halo with the mass and concentration of Antlia \citep{NFW1996,Jaffe2015}, while the solid black line marks the boundary of the virialised region following \citet{Mahajan2011MNRAS}. The galaxies in the grey area are inside the virialised region. The open grey squares represent galaxies associated with substructures. In each panel, the top and right subpanels show the corresponding 1D density distributions of $R/R_{200}$ and $|\Delta v|/\sigma$, respectively.}
\label{fig:PS_Q_SF_ETG_LTG}
\end{figure*}

Figure~\ref{fig:PS_Q_SF_ETG_LTG} shows the projected phase-space diagram for the Antlia spectroscopic sample. In panel~(a), galaxies are colour-coded by star-formation activity, with SFGs shown as cyan stars and QGs as magenta circles. In panel~(b), galaxies are colour-coded by morphology, with ETGs and LTGs represented by red squares and blue diamonds, respectively. In both cases, galaxies concentrate towards low $|\Delta v|/\sigma$ values, as expected for a dynamically bound population, with ETGs and QGs preferentially located at smaller projected radii. Kernel-density estimates in Appendix~\ref{sec:Appendix_PSD} provide a clearer view of these distributions.

Following \citet{Mahajan2011MNRAS}, we then separated the PPS into a high-probability virialised and non-virialised regions (infall and backsplash). Table~\ref{tab:galaxy_fractions_internal} shows the fractions of SFGs, QGs, LTGs, and ETGs in these two regimes, normalised to the total number of galaxies in each region. Quenched galaxies are more common in the virialised region, whereas SFGs and LTGs dominate the non-virialised region. This is consistent with the picture in which galaxies that have spent more time in the cluster potential are more likely to be quenched and morphologically transformed, while recently accreted systems retain late-type morphologies and ongoing star formation.

\begin{table}
\caption{Galaxy-type percentages in the PPS regions}
\label{tab:galaxy_fractions_internal}
\centering
\setlength{\tabcolsep}{4pt}
\begin{tabular}{p{3.cm}cccc}
\hline\hline
Zone & SFGs & QGs & LTGs & ETGs \\
\hline
Virialised      & $34\pm8$\% & $66\pm8$\% & $37\pm8$\% & $40\pm8$\% \\
Non-virialised  & $74\pm5$\% & $26\pm5$\% & $57\pm5$\% & $25\pm5$\% \\
\hline
\end{tabular}
\tablefoot{Fractions are completeness-corrected using weights $w$. The uncertainties correspond to $1\sigma$ approximate errors computed as $\sigma_p=\sqrt{p(1-p)/N_{\rm eff}}$, with $N_{\rm eff}=(\sum w)^2/\sum w^2$.}
\end{table}

\section{Discussion}\label{sec:discussion}

\subsection{Morphology--density relation and pre-processing}

In this work, we find that the spectroscopic sample analysed (galaxies brighter than $r_{\rm AUTO}=16$) is dominated by SFGs. These galaxies are present at all cluster-centric distances, but their fraction increases towards the outskirts, with many systems found between $3$ and $5\,R_{200}$. This behaviour is consistent with the well-known increase in SFG and LTG fractions with cluster-centric distance \citep{Dressler1980,Kauffmann2004MNRAS,Peng2010ApJ_quenching,Fasano2015,Haines2015ApJ,Liu2019,Lopes2024MNRAS}. The high fraction of star-forming systems at large radii, together with the presence of substructures in the outskirts, suggests that much of the galaxy population is still in an early stage of environmental processing.

Analysing the distribution of galaxy types across the virialised and non-virialised regions of the PPS (Table~\ref{tab:galaxy_fractions_internal}), we find clear trends with both morphology and star-formation activity. Inside the virialised region, QGs represent nearly two-thirds of the population ($66\pm8\%$), while SFGs account for only $34\pm8\%$. Outside this region, the situation reverses: SFGs dominate with $74\pm5\%$, and the fraction of quenched systems drops to $26\pm5\%$. This behaviour is consistent with a scenario in which galaxies gradually quench as they move from the infall-dominated, non-virialised region into the cluster core.

Morphologically, a similar segregation is observed. In the non-virialised region, LTGs dominate ($57\pm5\%$), and the ETG fraction decreases to $25\pm5\%$. In the virialised region, however, ETGs and LTGs have comparable fractions within the uncertainties, $40\pm8\%$ and $37\pm8\%$, respectively. Taken together with the SFG/QG trends, this suggests that the suppression of star formation may occur before the completion of morphological transformation, since the fraction of QGs in the virialised region is higher than that of ETGs.

It is important to emphasise that the \citet{Mahajan2011MNRAS} boundary was calibrated using cosmological simulations and should be interpreted statistically: virialised, infalling, and backsplash galaxies coexist in both regions \citep{Mahajan2011MNRAS,Rhee2017ApJ,Pasquali2019MNRAS}. Within these limitations, the significant population of QGs and ETGs found outside the virialised region indicates that some galaxies must have been at least partially quenched before reaching the cluster virialised region, most likely through environmental effects acting before they reached the cluster virialised region. This is consistent with pre-processing in group-scale haloes, although we cannot exclude the larger-scale overdense environment around Antlia, where quenching may also be linked to structures on megaparsec scales. These two scenarios are not mutually exclusive, since infalling groups are embedded in the surrounding large-scale structure. Conversely, the presence of SFGs and LTGs inside the virialised region shows that environmental quenching is not instantaneous and that a fraction of recently accreted systems can retain their gas and star formation even after completing at least one pericentric passage.

\subsection{Substructure in the Antlia cluster}\label{sec:substructures_D}

\texttt{CALSAGOS} is expected to identify substructures more reliably in the outer regions of clusters than in their inner regions. Tests presented by \citet{CALSAGOS2023MNRAS} show that the method performs best beyond $1R_{200}$. Nevertheless, \texttt{CALSAGOS} identifies a massive single substructure in the core of the Antlia cluster. However, this central substructure appears to consist of at least two galaxy groups in the process of interaction, in agreement with previous studies that also reported substructures in the central region \citep{Hess2015MNRAS,Calderon2020MNRAS}. For this reason, Table~\ref{tab:galaxy_types} presents the fractions of galaxy types both including and excluding the central massive substructure. Most of the other substructures identified by \texttt{CALSAGOS} consist of small galaxy groups located between $4$ and $5\,R_{200}$ (see Fig.~\ref{fig:substructure_Daniela}). These groups are likely recent infallers and are beginning to undergo pre-processing before fully entering the cluster environment
\citep{Haines2015ApJ,Lopes2024MNRAS}.

Table~\ref{tab:galaxy_types} summarises the distribution of galaxy types in different environments. If we exclude the central massive structure, we find that substructures in Antlia are dominated by SFGs, with 76$\pm$7\% classified as SFGs and 24$\pm$7\% as QGs. Including the central massive structure changes the balance to 47$\pm$6\%  QGs and 53$\pm$6\% SFGs, illustrating how the central substructure significantly increases the quenched fraction. This is in line with expectations: galaxies in the central substructure are likely to have resided longer in the dense cluster environment and to have experienced
processes such as ram-pressure stripping and tidal interactions, which are effective in quenching star formation.

The morphological mix follows a similar pattern. Excluding the central massive structure, substructures are dominated by LTGs, which comprise 60$\pm$8\% of the population, while ETGs account for 22$\pm$7\%. When the central structure is included, the ETG fraction increases to 31$\pm$5\%, although LTGs still remain more numerous (46$\pm$6\%). This again suggests that galaxies in the central substructure are more morphologically evolved, as expected for systems that have been subject to strong environmental effects for a longer time. An additional interesting result is that Antlia galaxies not associated with any substructure exhibit high fractions of SFGs ($78\pm$6\%).

\subsection{Comparison with the Hydra cluster}

In this study, we used the $(u-r)$ colour and S\'ersic index $n_{r}$ to classify galaxies as ETGs and LTGs \citep{vika2015A&A}. For the full spectroscopic sample out to $5\,R_{200}$, we find that 52$\pm$5\% of the galaxies are classified as LTGs, while 29$\pm$4\% are identified as ETGs. The remaining $\sim 19\%$ of galaxies occupy different regions of the $(u-r)$--$n_{r}$ plane. This high fraction of LTGs relative to ETGs is expected, since our analysis extends to large radii where infalling galaxies are more common and are expected to be predominantly LTGs.

The same methodology for classifying ETGs, LTGs, QGs, and SFGs using S-PLUS data has already been applied to the Hydra cluster by \citet{Lima-Dias2021MNRAS}, also for galaxies brighter than 16 in the $r$ band. A direct comparison therefore requires restricting Antlia to the same radial range. Considering only galaxies within $1\,R_{200}$, Antlia exhibits a similar fraction of ETGs and LTGs, 39$\pm$6\% and 38$\pm$6\%, respectively. In Hydra, by contrast, 54\% of galaxies are ETGs and only 23\% are LTGs \citep{Lima-Dias2021MNRAS}. Hydra is known to be a nearly relaxed cluster, with no strong evidence of recent interactions \citep{Fitchett1988,Hayakawa2004,Lima-Dias2021MNRAS}, whereas Antlia shows significant substructure, including two interacting groups \citep{Hess2015MNRAS,Calderon2020MNRAS}. Our results therefore suggest that the dynamical state of Antlia plays a crucial role in maintaining a higher fraction of LTGs compared to Hydra.

Cluster mass is another important difference between Antlia and Hydra: Hydra is more massive \citep[$M_{200} \simeq 2.2\times10^{14},M_{\odot}$,][]{Sifon2025A&A} than Antlia \citep[$M_{200} \simeq 1.3\times10^{14},M_{\odot}$,][]{Sifon2025A&A}. One might therefore expect Hydra to host a higher fraction of ETGs as a result of stronger environmental processing. We note, however, that the dependence of morphological fractions on cluster mass is not necessarily the same as the dependence of the passive fraction on cluster mass. While quenching is known to correlate strongly with both stellar mass and environment \citep[e.g.][]{Peng2010ApJ_quenching,Sobral2022ApJ}, morphological transformation can proceed on different timescales. Indeed, several observational studies of local clusters do not find a strong correlation between ETG fraction and cluster mass or velocity dispersion \citep{Postman2005ApJ,Poggianti2009ApJ,Simard2009A&A,Hoyle2012MNRAS}. This may indicate that, in the cluster regime, the morphology-environment relation is partly saturated, so that residual differences among clusters are more strongly affected by local density, accretion history, and dynamical state than by halo mass alone. Nevertheless, there are counter-examples: \citet{Oh2018ApJS}, analysing 14 clusters at $0.015 < z < 0.144$, find that more massive clusters host higher ETG fractions, and \citet{Pallero2026arXiv260619461P}, using the C-EAGLE simulations \citep{Barnes2017MNRA}, also find that more massive haloes have larger ETG fractions. In this sense, our Antlia-Hydra comparison is consistent with a picture in which both cluster mass and dynamical state contribute to the observed morphological mix.

We also find that Antlia hosts a significant fraction of SFGs: 65$\pm$4\% in the full sample are classified as star-forming. Within $\sim 1\,R_{200}$, \citet{Cairns2019ApJ} showed that Antlia members have molecular gas reservoirs comparable to field galaxies of similar stellar mass and SFR, contrary to what is observed in virialised clusters. A high fraction of SFGs in Antlia is therefore expected.

Restricting again to $R \leq 1\,R_{200}$, we find that 70$\pm$7\% of Antlia galaxies are quenched, while the remaining 30$\pm$7\% are still star-forming. In Hydra, \citet{Lima-Dias2021MNRAS} reported that $88\%$ of galaxies are quenched, a higher fraction than in Antlia. This contrast in SFG fraction may reflect a combination of cluster mass, assembly history, and dynamical state. In this context, the high molecular gas content reported by \citet{Cairns2019ApJ} may be related, at least in part, to Antlia's dynamically active state. Observationally, \citet{Stroe2017MNRAS} report that disturbed and merging clusters host a higher fraction of H$\alpha$ emitters than relaxed systems. Hydrodynamical simulations support this: \citet{Aldas2025A&A} report that disturbed groups and clusters contain higher fractions of blue and SFGs, as well as larger gas reservoirs, than relaxed systems. They also show that cluster interactions can produce an initial enhancement of star formation, followed by suppression as the system evolves towards equilibrium. This suggests that the elevated gas reservoirs and star-forming population in Antlia may be partly transient signatures of ongoing assembly. However, we cannot determine from the present data alone whether these properties are purely transient or whether they also reflect more structural differences, such as Antlia's lower mass, accretion history, or large-scale environment. We therefore interpret Antlia as a dynamically active, still-assembling system, consistent with previous evidence that it is less dynamically relaxed than Hydra \citep{Hu2023ApJ}.

Although a direct comparison with other nearby clusters such as Virgo and Fornax is not straightforward due to differences in methodology and datasets, some qualitative parallels can be drawn. Antlia appears to share similarities with Virgo, where quenching is more efficient in the core while star formation persists in the outskirts, consistent with recent infall \citep{Boselli2014A&A,Boselli2020A&A}. Fornax, in contrast, is a lower-mass cluster \citep{Sifon2025A&A} but has a more relaxed, dense core, with a marked deficit of SFGs and evidence of pre-processing in the infalling Fornax~A group \citep{Scott2014MNRAS,Iodice2019A&A,Janz2021A&A}. This comparison suggests that Antlia is likely a younger system than both Virgo and Fornax, though a more conclusive statement would require homogeneous analyses of these clusters using similar datasets and classification methods. We will be able to do this once the CHANCES survey begins.

\subsubsection{Caveats and future prospects}

The interpretation of our findings must be approached with caution. The spectroscopic redshifts used in this study were mainly obtained from a compilation of Southern Hemisphere observations and were not specifically designed for cluster-environment studies. Our goal here is therefore to make the best use of the existing spectroscopic information in the Antlia region, in order to obtain an overview of the distribution of different galaxy types, rather than to build a statistically complete census.

When examining the galaxy types in the identified substructures, we find that almost all of them are predominantly composed of SFGs. As discussed above, Antlia is a dynamically active cluster, and a relatively high fraction of SFGs is indeed expected even inside the cluster (up to $\sim 2R_{200}$). However, it is less plausible that nearly all substructures in the outskirts, particularly around $\sim 5R_{200}$, would consist exclusively of SFGs. Previous studies have shown that many galaxies undergo pre-processing in groups (i.e. small-scale structures) before falling into the cluster environment \citep[e.g.][]{Fujita2004PASJ,Pallero2019MNRAS}, so we would also expect to find a non-negligible population of QGs within these substructures. All these points suggest that our spectroscopic sample, at least in the outermost regions of the cluster, is likely biased towards star-forming systems.

These caveats do not invalidate our conclusions: all galaxies in our sample are genuine spectroscopic objects in the Antlia field. However, they highlight that our view of the galaxy population, especially in the outskirts, is incomplete and probably SFG-biased. This underlines the importance of dedicated surveys such as CHANCES for obtaining a more complete and homogeneous census of cluster galaxies. The analysis presented here will be revisited and refined once the full 4MOST/CHANCES data become available in the next 5 years of the survey.

\section{Summary and conclusions} \label{sec:conclusions}

In this work, we presented, for the first time, a structural analysis of galaxies in the Antlia cluster out to a projected radius of $5R_{200}$. Our main goal was to investigate how the environment influences galaxy properties. Our analysis focused on 154 spectroscopically confirmed cluster galaxies brighter than $r_{\rm AUTO}=16$. The imaging data were obtained from the S-PLUS survey, which provides photometry in 12 optical bands covering the visible spectrum. Galaxies were modelled with single Sérsic profiles using GALFITM and classified as ETGs and LTGs based on their Sérsic indices and colours (see Sect. \ref{sec:morph_class}). Our main results are summarised below. 

(i) Considering the cluster and its infall regions out to $5R_{200}$, Antlia hosts a higher fraction of LTGs (52$\pm$5\%) than ETGs (29$\pm$4\%). Within $1R_{200}$, however, the fractions of ETG and LTG are comparable considering the uncertainties. As a function of projected local density, LTGs dominate over ETGs at $\log(\Sigma_{10}\,[\mathrm{Mpc}^{-2}]) \lesssim 1.0$, while at higher densities the two populations become comparable within the uncertainties (see Fig.~\ref{fig:frac_Q_SF_LT_ET}). 

(ii) Antlia and its infall regions also contain a higher fraction of SFGs (65$\pm$4\%) compared to QGs when galaxies are considered out to $5R_{200}$. As a function of cluster-centric distance, QGs dominate over SFGs only within $1R_{200}$. When projected local density is used instead of radius, QGs become more numerous than SFGs only for $\log(\Sigma_{10}\,[\mathrm{Mpc}^{-2}]) \gtrsim 1.5$, consistent with the QGs/SFs–density relation (see Fig.~\ref{fig:frac_Q_SF_LT_ET}).

(iii) The PPS diagram shows a clear trend of segregation between the virialised and non-virialised regions defined following \citet{Mahajan2011MNRAS}. The virialised region is dominated by QGs, whereas the non-virialised region is dominated by SFGs and LTGs (Table~\ref{tab:galaxy_fractions_internal}). Nevertheless, the virialised region still contains a non-negligible fraction of SFGs and LTGs (of order 30–35\%), indicating that a significant number of systems have only recently reached the cluster core and are still undergoing quenching. This reinforces the picture in which Antlia is actively accreting galaxy groups and field galaxies from its surroundings, many of which are only beginning to experience strong environmental processing as they move towards the cluster centre (see Fig.~\ref{fig:PS_Q_SF_ETG_LTG}).

(iv) We identified substructures in Antlia out to $5R_{200}$. Almost all substructures beyond $1R_{200}$ are dominated by SFGs. We also find that galaxies not associated with any detected substructure exhibit a higher fraction of SFGs (78$\pm$6\%) and LTGs (60$\pm$7\%). However, when the central massive substructure is included, the QG and ETG fractions increase to $47\pm6\%$ and $31\pm5\%$, respectively. This suggests that the central substructure hosts a more environmentally processed population, while the outer substructures are dominated by star-forming and late-type systems.

In summary, Antlia is dominated by SFGs and LTGs between $1$ and $5R_{200}$, while showing a high fraction of QGs within $1R_{200}$. Together with the presence of multiple substructures, these results support the picture of Antlia as a dynamically active and young cluster that is still assembling its galaxy population. Figures~\ref{fig:substructure_Daniela} and \ref{fig:PS_Q_SF_ETG_LTG} clearly illustrate this dynamic nature of Antlia.

Taken together, these results indicate that Antlia is an example of ongoing cluster assembly. The persistence of SFGs within the virial region, together with the different galaxy populations found in the central and outer substructures, points to a scenario in which galaxy evolution is neither instantaneous nor driven by a single mechanism but rather reflects a continuous interplay between group, cluster, and larger scale environments. Antlia therefore provides a valuable benchmark for understanding how environmental processes operate in the local Universe. Future analyses within the CHANCES survey will place these findings in a statistical context, allowing us to explore how such processes depend on cluster mass, dynamical state, and accretion history. In particular, the forthcoming spectroscopic data will provide much higher completeness down to the dwarf regime, enabling a robust study of Antlia, using a homogeneously derived spectroscopic sample, and a direct comparison with a representative sample of nearby clusters.

\bibliographystyle{aa} 
\bibliography{references}

\FloatBarrier 
\twocolumn

\begin{appendix}
\section{Acknowledgments}
\begin{acknowledgements}
We thank the anonymous referee for their careful reading of the manuscript and for their constructive comments, which helped improve the clarity and quality of this work. CL-D and AM acknowledge a grant from the ESO Comité Mixto ORP037/2022, and the support from the Agencia Nacional de Investigación y Desarrollo (ANID) through Fondecyt project 3250511. AM acknowledges support from the ANID FONDECYT Regular grant 1251882, from the ANID BASAL project FB210003, and funding from the HORIZON-MSCA-2021-SE-01 Research and Innovation Programme under the Marie Sklodowska-Curie grant agreement number 101086388. G.M. gratefully acknowledges the Fundação de Amparo à Pesquisa do Estado de São Paulo (FAPESP) for the support grant 2024/10923-3. HMH gratefully acknowledges support from Agencia Nacional de Investigación y Desarrollo (ANID) through
Fondecyt project 3230176, ANID MILENIO NCN2024\_112, Fondo Rubin-Chile 2024, DIA2322 and ANID BASAL project FB210003. A.R.L. acknowledges the grant 2025/09544-0 from São Paulo Research Foundation (FAPESP). R.F.H and A.R.L. acknowledge financial support from Consejo Nacional de Investigaciones Cientificas y Técnicas (CONICET) (PIP 1504), Agencia I+D+i (PICT 2019–03299) and Universidad Nacional de La Plata (Argentina). YLJ acknowledges support from the Agencia Nacional de Investigaci\'on y Desarrollo (ANID) through Basal project FB210003, FONDECYT Regular projects 1241426 and 123044, and  Millennium  Science Initiative Program NCN2024\_112. STF acknowledges the financial support of DIDULS/ULS through a regular project number PR2453858 and the funding ADI2553855. AAC acknowledges financial support from the Severo Ochoa grant CEX2021-001131-S funded by MCIN/AEI/10.13039/501100011033 and the project PID2023-153123NB-I00 funded by MCIN/AEI. Part of this work was supported by the Polish Ministry of Science and Higher Education (MNiSW) grant DIR/WK/2018/12. A.C.K. thanks the Fundação de Amparo à Pesquisa do Estado de São Paulo (FAPESP) for the support grant 2024/05467-9 and the Conselho Nacional de Desenvolvimento Científico e Tecnológico (CNPq) grant 315566/2023-0. M.S.C acknowledges funding from São Paulo Research Foundation (FAPESP) grant 2023/10774-5 and funding to the S-PLUS project through the process 2019/26492-3. R.D. gratefully acknowledges support by the ANID BASAL project FB210003. E.I. gratefully acknowledge financial support from ANID - MILENIO - NCN2024\_112 and ANID FONDECYT Regular 1221846. SW is supported by the United Kingdom Research and Innovation (UKRI) Future Leaders Fellowship `Using Cosmic Beasts to uncover the Nature of Dark Matter' (grant number MR/X006069/1). UK acknowledges financial support from the UK Science and Technology Facilities Council (STFC; grant ref: ST/T000171/1). LSJ acknowledges the support from CNPq (308994/2021-3) and FAPESP (2011/51680-6). AC acknowledges the Fundação de Amparo à Pesquisa do Estado do Rio de Janeiro (FAPERJ) grant E26/202.607/2022 e 210.371/2022(270993) and the Bolsa de Produtividade do CNPQ level 2. EVRL acknowledges the financial support given by CAPES (88887.470064/2019-00), CNPq (169181/2017-0), and FAPESP (2024/15229-8). CPH acknowledges support from ANID through Fondecyt Regular project number 1252233. MAF acknowledges support from the Emergia program (EMERGIA20\_38888) from Junta de Andalucía and University of Granada. F.R.H. acknowledges support from FAPESP grants 2018/21661-9 and 2021/11345-5. F.A.-F. acknowledges support from FAPESP grants 2024/00822-5 and 2024/22842-8. D.E.O-R. acknowledges financial support from ANID Fondo ALMA 2025 project 31250033. IL acknowledges support from the ANID FONDECYT Regular grant 1261197. BFR acknowledges support for this publication from the Polish Ministry of Science and Higher Education (MNiSW) grants DIR/WK/2018/12 and 2026/WK/06.

The S-PLUS project, including the T80-South robotic telescope and the S-PLUS scientific survey, was founded as a partnership between the Fundação de Amparo à Pesquisa do Estado de São Paulo (FAPESP), the Observatório Nacional (ON), the Federal University of Sergipe (UFS), and the Federal University of Santa Catarina (UFSC), with important financial and practical contributions from other collaborating institutes in Brazil, Chile (Universidad de La Serena), and Spain (Centro de Estudios de Física del Cosmos de Aragón, CEFCA). We further acknowledge financial support from the São Paulo Research Foundation (FAPESP), the Brazilian National Research Council (CNPq), the Coordination for the Improvement of Higher Education Personnel (CAPES), the Carlos Chagas Filho Rio de Janeiro State Research Foundation (FAPERJ), and the Brazilian Innovation Agency (FINEP).

The authors are grateful for the contributions of CTIO staff in helping in the construction, commissioning and maintenance of the T80-South telescope and camera. We are also indebted to Rene Laporte and INPE, as well as Keith Taylor, for their important contributions to the project. We also thank CEFCA staff for their help with T80-South. Specifically, we thank Antonio Marín-Franch for his invaluable contributions in the early phases of the project, David Cristóbal-Hornillos and his team for their help with the installation of the data reduction package jype version 0.9.9, César Íñiguez for providing 2D measurements of the filter transmissions, and all other staff members for their support.

\end{acknowledgements}

\section{CHANCES Low-$z$ Antlia targeting strategy}
\label{app:chances}

For 4MOST operations, the CHANCES Low-$z$ target catalogue around Antlia is split into three target classes, following the general strategy adopted for the full Low-$z$ sample \citep{MendezHernandez2026A&A}. All three lists are drawn from the parent photometric sample defined in Sect.~\ref{sec:CHANCES_Members} (24\,751 extended sources within $5R_{200}$ and $r_{\rm AUTO} < 20.4$).

\begin{itemize}
  \item S1501 – Low-$z$ Bright. This sub-survey contains all galaxies brighter than $r_{\rm AUTO} = 18.5$. For Antlia, after visual inspection to remove obvious contaminants (e.g. stars, diffraction spikes, image artefacts), the S1501 list comprises 7\,145 targets. These objects have the highest fibre-allocation priority within CHANCES Low-$z$.

  \item S1505 – Low-$z$ Faint. This sub-survey includes galaxies fainter than $r_{\rm AUTO} = 18.5$ that satisfy the colour–magnitude selection described Sect.~\ref{sec:CHANCES_Members} and have either S-PLUS photometric redshifts $z_{\rm phot} < 0.2$ or no photometric redshift available. Galaxies without $z_{\rm phot}$ may still be bona fide cluster members, and are therefore retained to maximise the completeness of potential Antlia members in the main faint catalogue.

  \item S1506 – Low-$z$ Faint Supplementary. This supplementary list is built to optimise fibre usage at faint magnitudes. It contains all S1505 targets plus additional galaxies with $r_{\rm AUTO} > 18.5$ and $z_{\rm phot} > 0.2$, which are expected to be predominantly background systems. These objects have lower priority than S1501 and S1505.
\end{itemize}

The numbers above are relevant for the operational planning of the CHANCES Antlia observations, but they are not used in the scientific analysis presented in this paper, which is based exclusively on the
spectroscopically confirmed Antlia members (Sect.~\ref{sec:spec_Members}).

\section{Spectroscopic completeness and weighting}
\label{sec:appendix_completeness}

\begin{figure}
    \centering
    \includegraphics[width=\columnwidth]{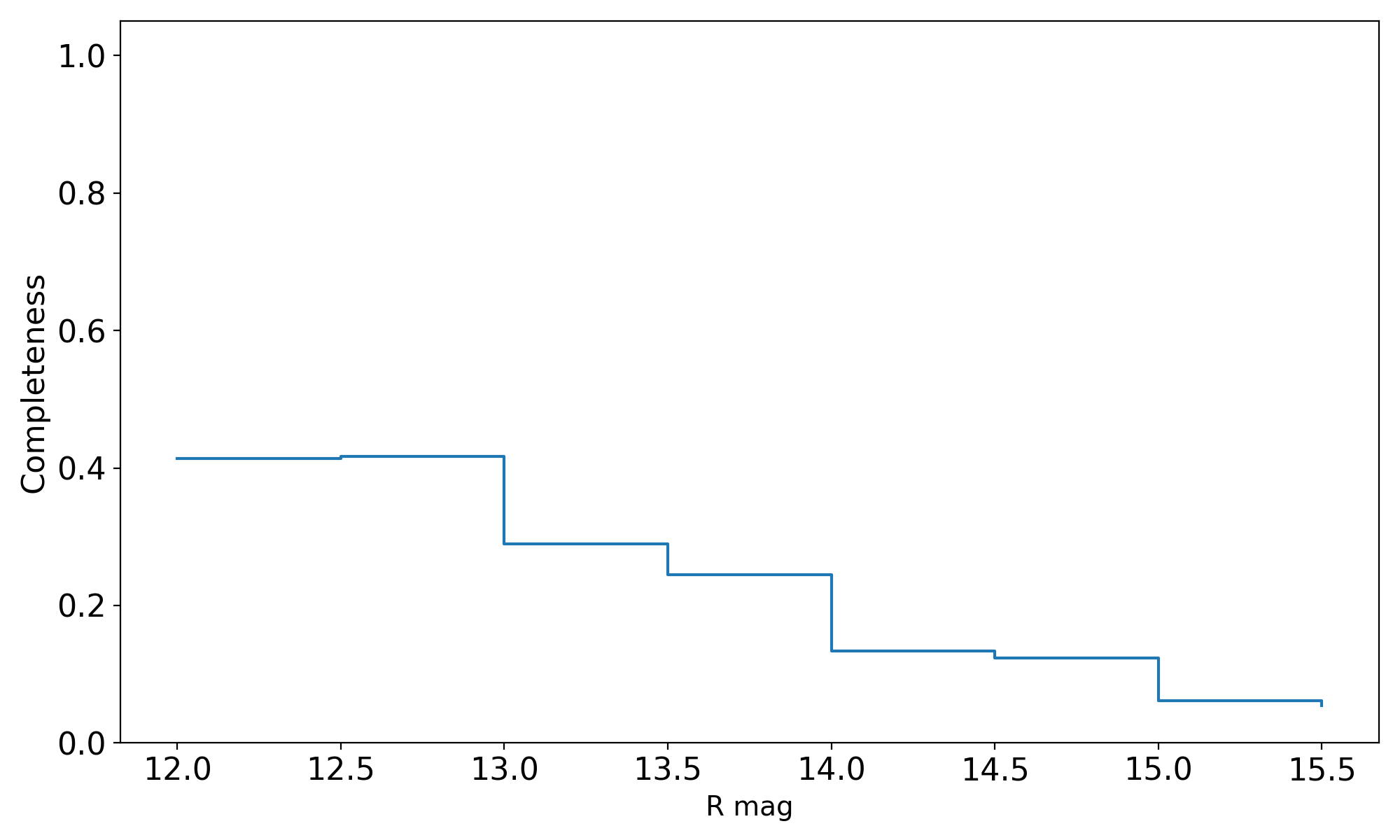}
    \includegraphics[width=\columnwidth]{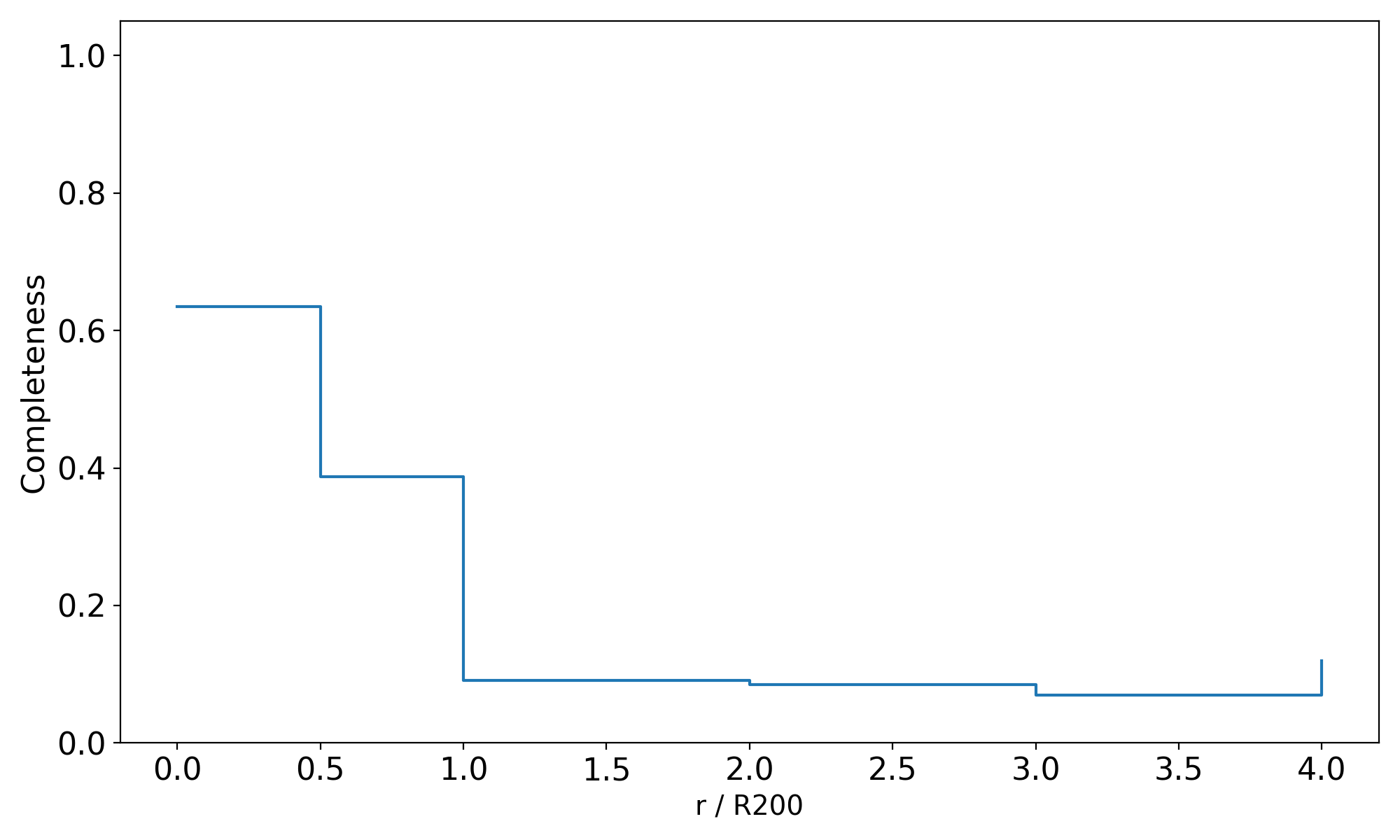}
    \caption{Spectroscopic completeness for galaxies brighter than
    $r_{\rm AUTO}=16$ within $5R_{200}$ of Antlia. Top: Completeness
    as a function of $r$-band magnitude, defined as the ratio between the
    number of galaxies with reliable redshifts and the total number of
    photometric detections in each magnitude bin. Bottom: Same as in the top panel but
    as a function of the projected cluster-centric distance expressed in units of $R_{200}$.
    These completeness estimates were used to define the weights applied in
    the analysis of galaxy fractions.}
    \label{fig:completeness}
\end{figure}

To quantify the impact of the heterogeneous spectroscopic coverage, we
estimated the completeness as a function of apparent $r$-band magnitude
and projected cluster-centric distance for all galaxies brighter than
$r_{\rm AUTO}=16$ within $5R_{200}$ of Antlia. In each bin of magnitude
and radius, we computed the ratio $C = N_{\rm spec}/N_{\rm phot}$ between
the number of galaxies with reliable redshifts and the total number of
photometric detections. The two panels of Fig.~\ref{fig:completeness} show
the resulting 1D completeness curves as a function of
$r$-band magnitude and $R/R_{200}$. These completeness estimates are used to define the weights $w = 1/C$ applied when computing galaxy fractions as a function of environment.

\section{Stellar-mass distribution of the spectroscopic sample}
\label{sec:mass_appendix}

\begin{figure}
    \centering
    \includegraphics[width=\columnwidth]{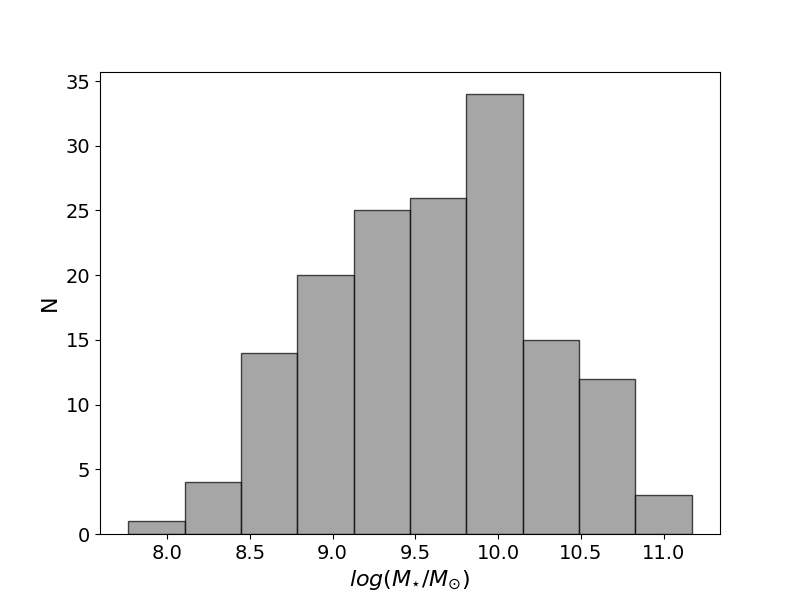}
    \caption{Stellar-mass distribution of the final spectroscopic sample.
    The histogram shows the number of galaxies as a function of 
    $\log(M_\star/M_\odot)$, with masses estimated using the 
    \citet{Taylor2011MNRAS} colour relation (Sect.~\ref{sec:stellar_mass}).}
    \label{fig:hist_mass}
\end{figure}

Figure~\ref{fig:hist_mass} shows the distribution of stellar masses for the 
154 galaxies in our final spectroscopic sample (Sect.~\ref{sec:spec_Members}). 
The sample spans 
$\log(M_\star/M_\odot) \approx 8.0$--$11.1$, with a median 
$\log(M_\star/M_\odot) \approx 9.7$. 
This confirms that our analysis is primarily sensitive to intermediate- and 
high-mass galaxies in Antlia.

\section{Excess emission in the $J0660$ filter: the three–filter method (3FM)}
\label{sec:3FM}

To classify galaxies as line emitters, we require that the \Ha\ equivalent width measured in the $J0660$ filter, $EW_{J0660}$, exceeds $12~\text{\AA}$. This threshold follows \citet{Vilella-Rojo2015A&A}, who showed that J-PLUS cannot robustly measure $EW_{J0660}$ values below $12~\text{\AA}$ at the $3\sigma$ level. Given that S-PLUS and J-PLUS are twin facilities with identical filter systems, we adopt the same criterion in this work. The equivalent width in $J0660$ is computed as

\begin{equation}
   EW_{J0660} = \Delta_{J0660}\,\frac{Q - 1}{1 - Q\,\epsilon},
	\label{eq:equivalent_with}
\end{equation}

where $EW_{J0660}$ is the equivalent width in the $J0660$ band, $\Delta_{J0660}$ is the width of the $J0660$ filter, $Q$ is defined by
$m_{r} - m_{J0660} = 2.5 \log Q$, and $\epsilon \equiv \Delta_{J0660} / \Delta_{r}$.

\begin{figure}
\centering
\includegraphics[width=\columnwidth]{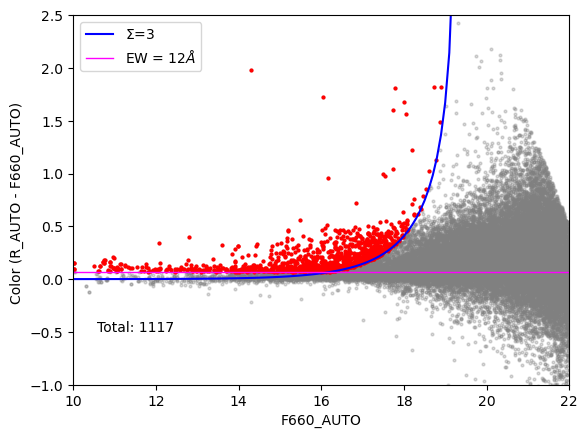}
    \caption{Application of 3FM to the S-PLUS field HYDRA-0052. The $J0660$ excess, $m_{r} - m_{J0660}$, plotted as a function of $m_{J0660}$. The blue curve shows the $3\Sigma$ significance threshold, and the horizontal magenta line marks the $EW_{J0660} = 12~\text{\AA}$ cut. The grey points represent all detected sources, while the red points indicate objects that satisfy both criteria and are therefore classified as significant narrow-band emitters.}
    \label{fig:3FM}
\end{figure}

In addition, we require a significant colour excess relative to the expected scatter of sources with zero intrinsic colour. This is quantified by the $\Sigma$ parameter and a $3\Sigma$ cut \citep{Sobral2012MNRAS}. The corresponding curve is given by \citet{Khostovan2020MNRAS} as

\begin{equation}
   \Sigma = 1 - 
   \frac{10^{-0.4\,(m_{r} - m_{J0660})}}
        {10^{ZP - m_{J0660}}\,
         \sqrt{\sigma_{J0660}^{2} + \sigma_{r}^{2}} } ,
	\label{eq:sigma_cut}
\end{equation}

where $m_{r}$ and $m_{J0660}$ are the magnitudes in the $r$ and $J0660$ bands, respectively; $ZP$ is the photometric zero–point of the $J0660$ image; and $\sigma_{J0660}$ and $\sigma_{r}$ are the corresponding magnitude uncertainties. Figure~\ref{fig:3FM} illustrates this selection for one S-PLUS field: grey points show all detected sources, while red points highlight those that satisfy both the $EW_{J0660} > 12~\text{\AA}$ and $\Sigma \ge 3$ criteria. The blue curve and magenta line mark the $3\Sigma$ threshold and the $EW_{J0660}=12~\text{\AA}$ cut, respectively.

Once galaxies with a genuine $J0660$-band excess have been identified, we use the three–filter method \citep[3FM;][]{Pascual2007PASP} to estimate the combined \Ha\,+\nii\ flux. The analytical expression of the 3FM is

\begin{equation}
F_{\mathrm{H}\alpha + [\mathrm{NII}]} = 
\frac{
    \left( \overline{F_{r}} - \overline{F_{i}} \right) 
    - \left( \frac{\alpha_{r} - \alpha_{i}}{\alpha_{J0660} - \alpha_{i}} \right) 
      \left( \overline{F_{J0660}} - \overline{F_{i}} \right)
}{
    \beta_{J0660} \left( \frac{\alpha_{i'} - \alpha_{r}}{\alpha_{J0660} - \alpha_{i}} \right) 
    + \beta_{r}
},
\label{eq:3FME}
\end{equation}

where $\overline{F_{\lambda}}$ denotes the mean flux density in each filter, and the coefficients $\alpha_{\lambda}$ and $\beta_{\lambda}$ are functions of the effective wavelengths and bandpasses of the filters (see \citealt{Vilella-Rojo2015A&A} for a full derivation).

\section{Non-parametric morphology classification}
\label{sec:nonparam_appendix}

In order to investigate possible bias in our analysis due to the morphological scheme adopted, we use non-parametric indexes in an analogue way to Fig.~\ref{fig:vika_density}. Namely, we use the python-package galaxy morphology extractor \citep[galmex,][]{Sampaio2026A&A} to measure Concentration (C), Asymmetry (A), Smoothness (S), M20, Shannon Entropy (E), Gini index (G), Gradient Pattern Asymmetry (G2) for S-PLUS r-band images. Prior to metrics estimation, the pipeline performs background subtraction, object detection, secondary objects cleaning, and segmentation mask creation. In this work, we adopt the procedure described in \cite{2024MNRAS.528...82K}, which follows from the similar pixel scale and point spread function with respect to the SDSS.

In brief, the CAS+MEGG definitions are the following: 1) the $C$ index is defined from the elliptical radii enclosing 20\% and 80\% of the total light, $R_{20}$ and $R_{80}$; 2) Asymmetry $A$ is computed by comparing the original image with it's 180 degrees rotated version; 3) smoothness $S$ compares the original image with a smoothed version, in order to filter high-frequencies; 4) M20 calculates the logarithmic ratio of the second order moment of the pixels summing up to 20\% of galaxy total flux, and the total second order momentum; 5) E is computed by creating a histogram of the pixel values and then calculating the Shannon entropy equation; 6) the Gini index is defined as the area between a Lorentzian curve fitting the galaxy pixel values distribution and the equality line (x=y); and 7) G2 is defined as the asymmetry of the pixel values gradient field when compared to its 180 degrees rotated version. See \cite{2024MNRAS.528...82K} for an extended description.

\begin{figure}
\centering
\includegraphics[width=\columnwidth]{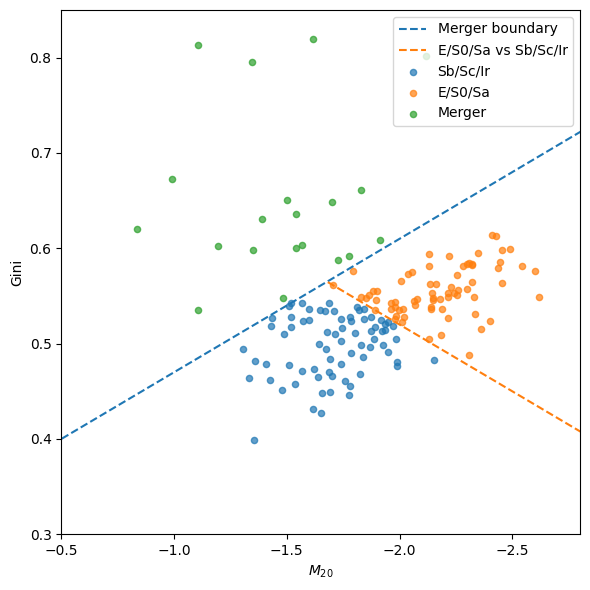}
\caption{Gini–$M_{20}$ diagram for Antlia galaxies in the spectroscopic sample. The symbols indicate the regions defined by
\citet{Lotz2008ApJ}: mergers (squares), late-type (Sb/Sc/Ir; stars),
and early-type (E/S0/Sa; circles). The dashed lines show the
corresponding empirical boundaries. The majority of galaxies
classified as ETGs from the $(u-r, n_r)$ plane lie in the E/S0/Sa
region, while LTGs populate mainly the Sb/Sc/Ir locus, with only a
few systems approaching the merger boundary.}
\label{fig:gini_m20_appendix}
\end{figure}

\begin{figure}
\centering
\includegraphics[width=\columnwidth]{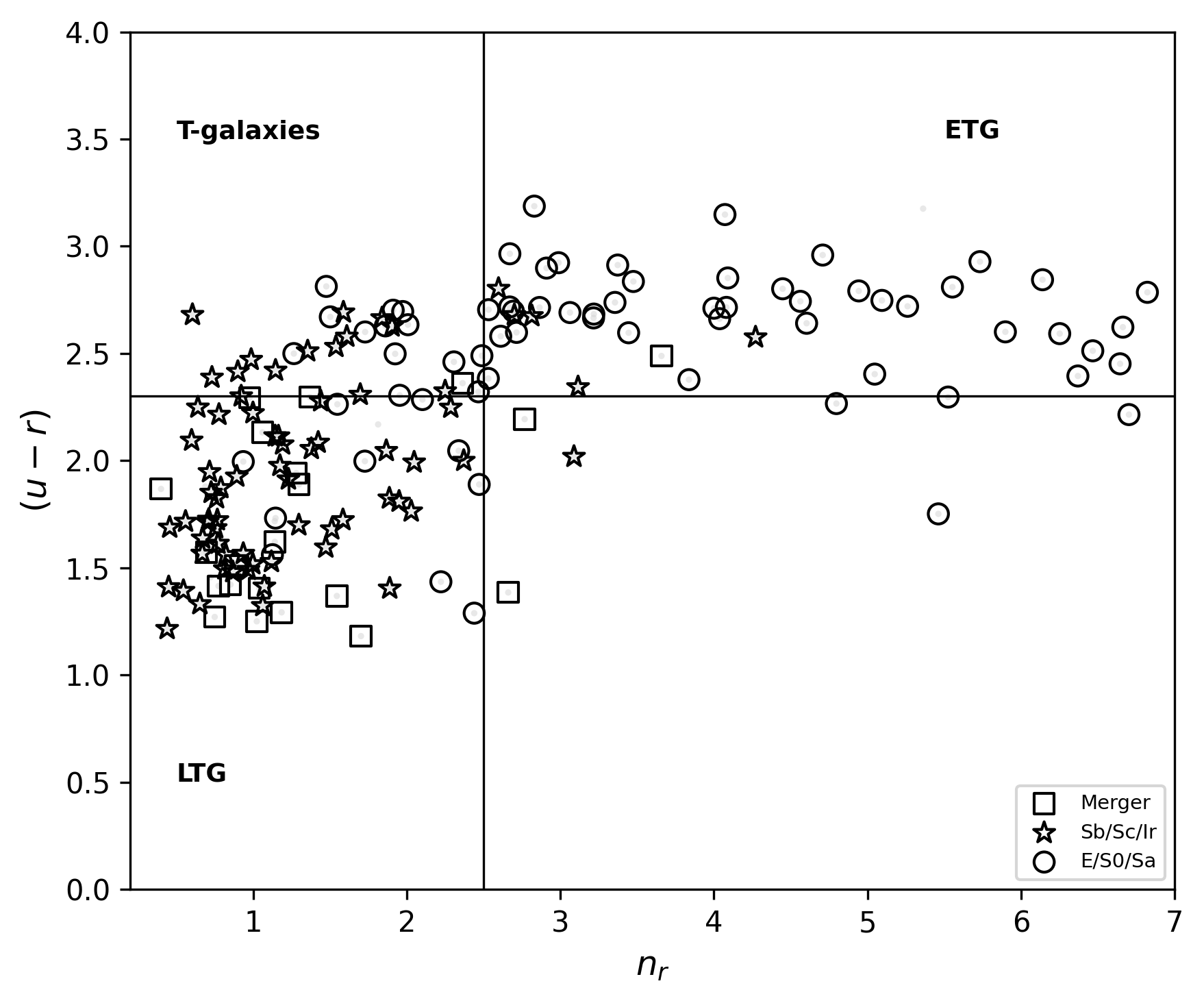}
\caption{Distribution of the \citet{Lotz2008ApJ} morphological classes in the $(u-r)$-$n_r$ plane for the Antlia spectroscopic sample. The open squares indicate mergers, the stars represent Sb/Sc/Ir systems, and the circles denote E/S0/Sa galaxies. The vertical and horizontal lines mark the adopted divisions at $n_r=2.5$ and $(u-r)=2.3$, separating the LTG, T-galaxy, and ETG regions defined in Sect.~\ref{sec:morph_class}.}
\label{fig:nonparam_ur_nr_appendix}
\end{figure}

In Fig.~\ref{fig:gini_m20_appendix} we present the Gini-$M_{20}$ diagram, together with the empirical boundaries defined by \citet{Lotz2008ApJ} to separate mergers, late-type (Sb/Sc/Irr), and early-type (E/S0/Sa) systems. In this diagram, the Gini coefficient quantifies how concentrated the galaxy light is among its pixels, while $M_{20}$ traces the spatial distribution of the brightest 20\% of the light. Early-type systems generally occupy the region of high Gini and low $M_{20}$, reflecting centrally concentrated light distributions, whereas late-type galaxies tend to have lower Gini values and higher $M_{20}$ due to more extended and structured light distributions. Merger candidates are typically offset from the normal galaxy sequence because multiple bright nuclei, tidal features, or strong asymmetries increase the relative contribution of bright off-centre structures. Figure~\ref{fig:nonparam_ur_nr_appendix} shows these \citet{Lotz2008ApJ} classifications projected onto the $(u-r)$--$n_r$ plane. We find that the ETG region is predominantly occupied by E/S0/Sa systems, whereas the LTG region is mainly populated by Sb/Sc/Ir systems and mergers.

\section{Relation between projected local density and cluster-centric distance}
\label{sec:density_radius}

\begin{figure}
\centering
\includegraphics[width=\columnwidth]{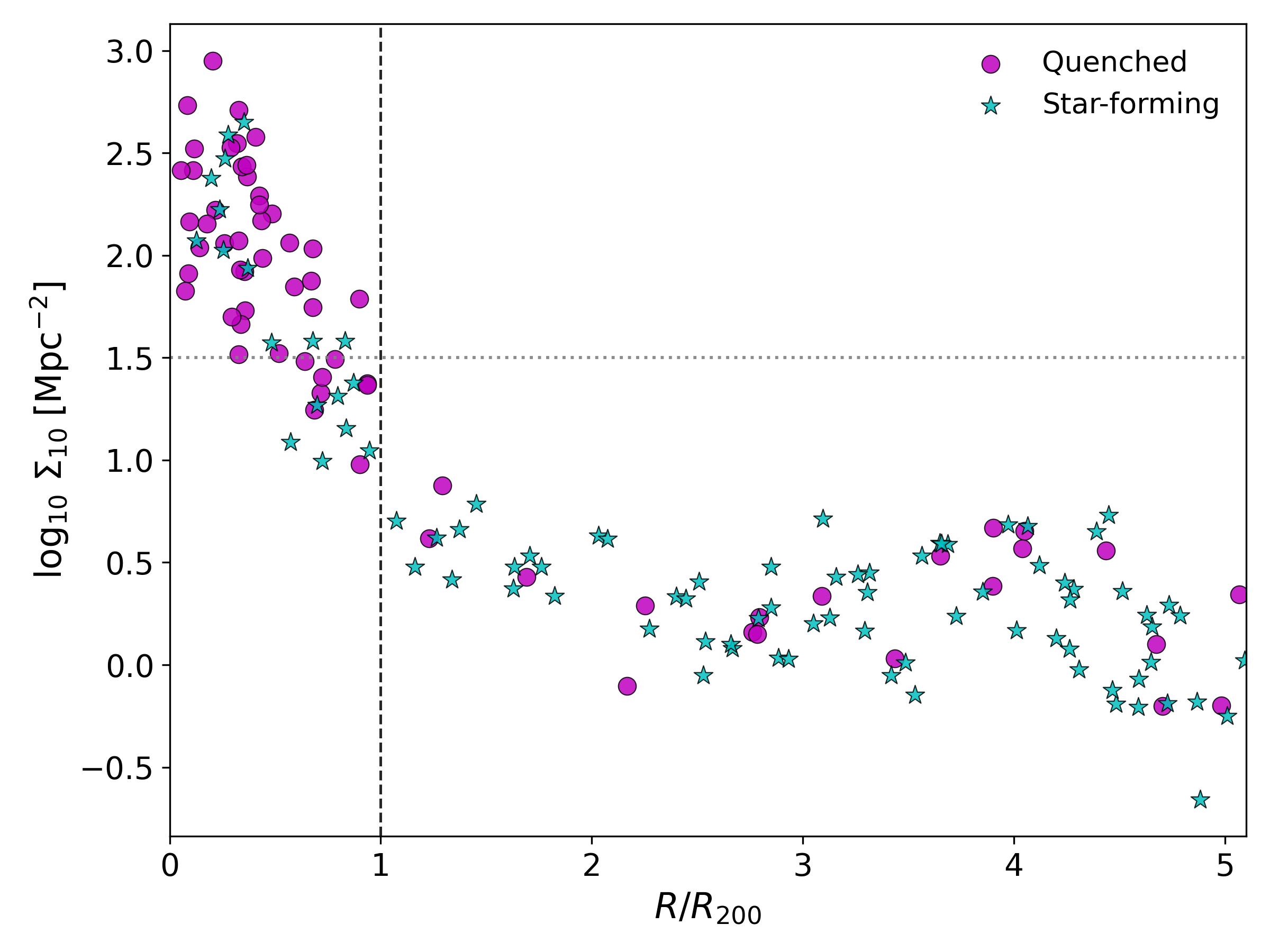}
\caption{Projected local density, $\log_{10}\Sigma_{10}$, as a function of projected cluster-centric distance, $R/R_{200}$, for QGs (magenta circles) and SFGs (cyan stars). The vertical dashed line marks $R_{200}$, and the horizontal dotted line indicates $\log_{10}\Sigma_{10}=1.5$, the approximate density above which QGs become more frequent than SFGs in Fig.~\ref{fig:frac_Q_SF_LT_ET}.}
\label{fig:density_radius_qsf}
\end{figure}

Figure~\ref{fig:density_radius_qsf} shows the relation between projected local density, $\log_{10}\Sigma_{10}$, and projected cluster-centric distance, $R/R_{200}$, for QGs and SFGs separately. The highest projected local densities, $\log_{10}\Sigma_{10}\gtrsim1.5$, are mainly found within $R_{200}$ and are mostly populated by QGs. Beyond $R_{200}$, both QGs and SFGs are generally located in lower-density environments, although SFGs dominate numerically at these radii. This confirms that $\Sigma_{10}$ and $R/R_{200}$ are partially correlated in our sample, and that part of the trend observed with local density in Fig.~\ref{fig:frac_Q_SF_LT_ET} reflects the underlying radial distribution of galaxy types.

\section{Projected phase–space diagrams}
\label{sec:Appendix_PSD}

\begin{figure}
\centering
\includegraphics[width=\columnwidth]{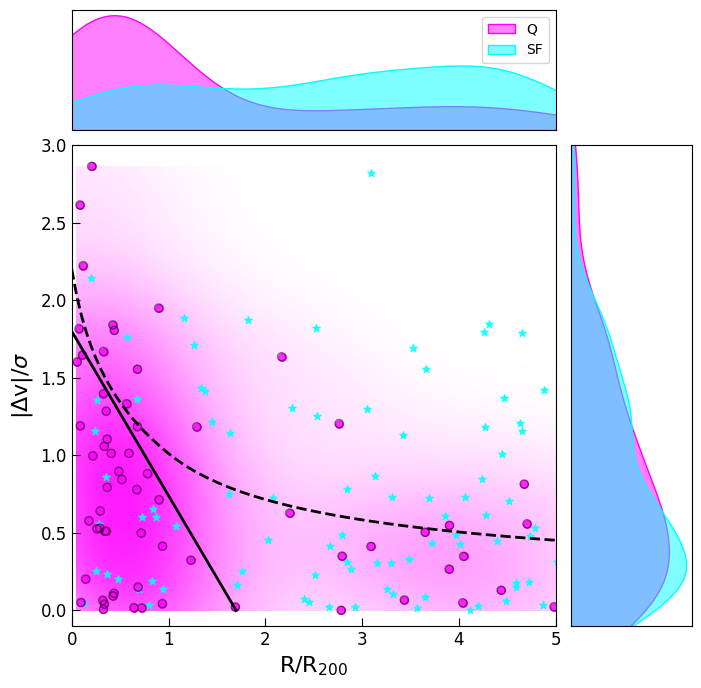}
\caption{PPS diagram as in
Fig.~\ref{fig:PS_Q_SF_ETG_LTG}. The cyan stars and magenta circles represent SFGs and
QGs, respectively. The background density map illustrates the distribution of
QGs.}
\label{fig:PS_quenched_KDE}
\end{figure}

\begin{figure}
\centering
\includegraphics[width=\columnwidth]{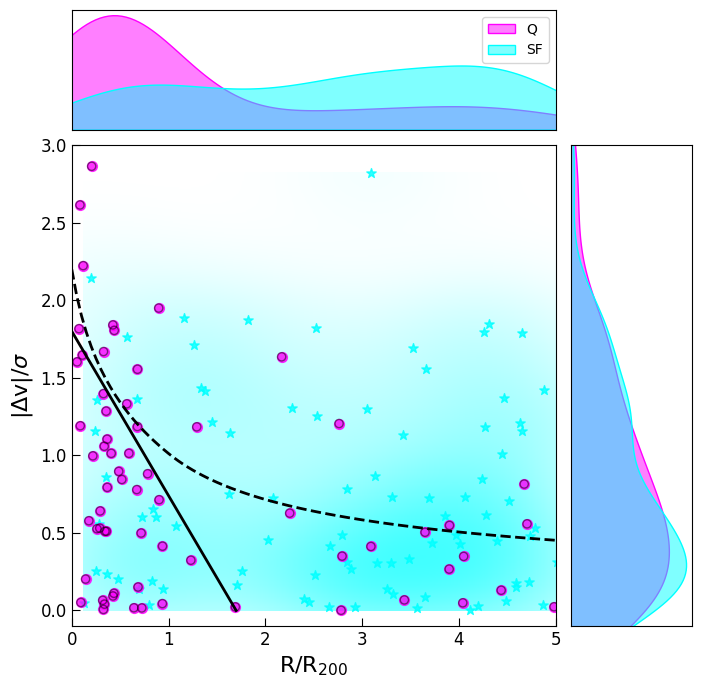}
\caption{PPS diagram as in
Fig.~\ref{fig:PS_Q_SF_ETG_LTG}. The cyan stars and magenta circles represent SFGs and
QGs, respectively. The background density map illustrates the distribution of
SFGs.}
\label{fig:PS_SF_KDE}
\end{figure}

\begin{figure}
\centering
\includegraphics[width=\columnwidth]{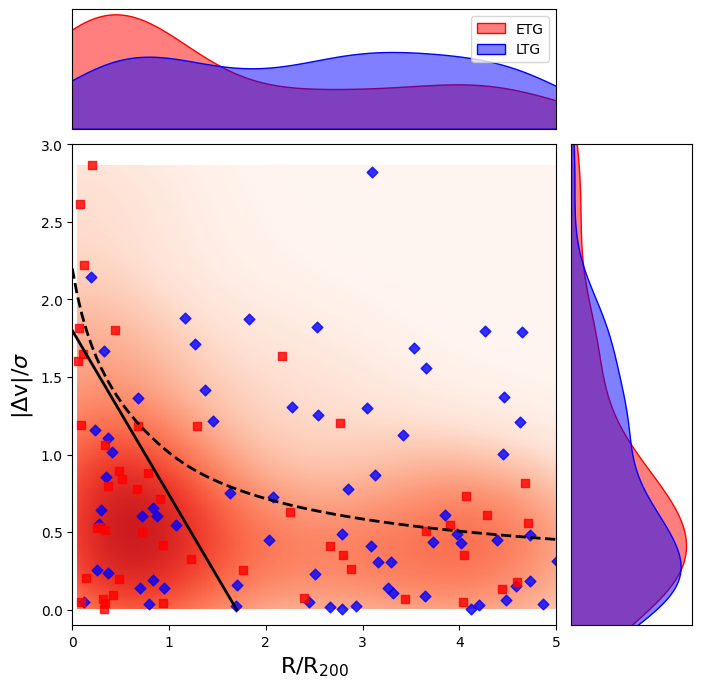}
\caption{PPS diagram as in
Fig.~\ref{fig:PS_Q_SF_ETG_LTG}. The blue diamonds and red squares represent LTGs and
ETGs, respectively. The background density map illustrates the distribution of
ETGs.}
\label{fig:PS_ETG_KDE}
\end{figure}

\begin{figure}
\centering
\includegraphics[width=\columnwidth]{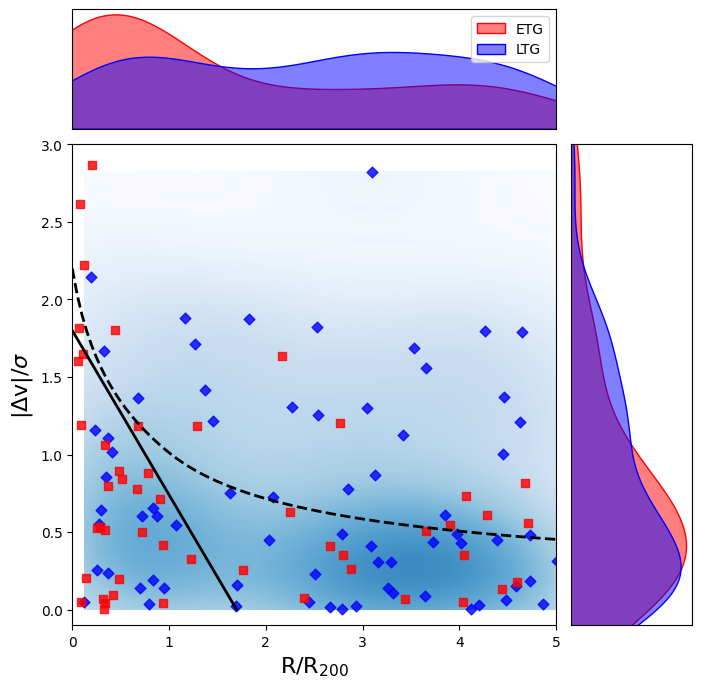}
\caption{PPS diagram as in
Fig.~\ref{fig:PS_Q_SF_ETG_LTG}. The blue diamonds and red squares represent LTGs and
ETGs, respectively. The background density map illustrates the distribution of
LTGs.}
\label{fig:PS_LTG_KDE}
\end{figure}

The PPS diagram is analysed in the main body of the paper (Sect.~\ref{sec:ps_section}). Here we provide complementary visualisations using 2D kernel–density estimates to highlight more clearly where different galaxy populations are concentrated in phase-space. Figures~\ref{fig:PS_quenched_KDE} and \ref{fig:PS_SF_KDE} show the distributions for QGs and SFGs, respectively, while Figs.~\ref{fig:PS_ETG_KDE} and \ref{fig:PS_LTG_KDE} show the corresponding distributions for early–type (ETGs) and late–type galaxies (LTGs).

\end{appendix}

\end{document}